\documentclass[12pt]{article}
\usepackage[utf8]{inputenc}
\usepackage{amsmath,epsfig,amssymb,bm,amsfonts,amsthm,epsfig,verbatim, enumitem}
\usepackage[numbers, ,sort&compress]{natbib}%[comma,authoryear]
\usepackage{color,graphicx,lscape,longtable,mathabx,natbib, float, multirow}
\usepackage{hyperref,subfigure, multirow, setspace,  booktabs}
\usepackage{algorithm}
\usepackage{algorithmic}
\usepackage{makecell}
\usepackage[nolists,figuresfirst,nomarkers]{endfloat} % nomarkers
\usepackage{ulem} % delete line

\graphicspath{{figures/}}

\newtheorem{Theorem}{Theorem}[section]

\def\bse{\begin{eqnarray*}}
\def\ese{\end{eqnarray*}}
\def\be{\begin{eqnarray}}
\def\ee{\end{eqnarray}}
\def\bsq{\begin{equation*}}
\def\esq{\end{equation*}}
\def\bq{\begin{equation}}
\def\eq{\end{equation}}

\def\wh{\widehat}

\def\trans{^{\rm T}}

\def\bmu{\boldsymbol\mu}

\def\bmu{\boldsymbol\mu}
\def\bg{\boldsymbol\gamma}

\def\btheta{\boldsymbol\theta}

\def\bbo{{\bf 0}}
\def\0{{\bf 0}}

\def\B{{\bf B}}

\def\f{{\bf f}}
\def\h{{\bf h}}

\def\b{{\bf b}}

\def\I{{\bf I}}

\def\F{{\bf F}}
\def\H{{\bf H}}

\def\X{{\bf X}}

\def\x{{\bf x}}
\def\I{{\bf I}}

\def\Y{{\bf Y}}

\def\bq{\begin{equation}}
\def\eq{\end{equation}}

\def\wh{\widehat}

\def\trans{^{\rm T}}

\def\squarebox#1{\hbox to #1{\hfill\vbox to #1{\vfill}}}

\def\bse{\begin{eqnarray*}}
\def\ese{\end{eqnarray*}}
\def\be{\begin{eqnarray}}
\def\ee{\end{eqnarray}}
\def\bsq{\begin{equation*}}
\def\esq{\end{equation*}}
\def\bq{\begin{equation}}
\def\eq{\end{equation}}
\def\wh{\widehat}

\def\trans{^\mathrm{T}}

\def\boxit#1{\vbox{\hrule\hbox{\vrule\kern6pt\vbox{\kern6pt#1\kern6pt}\kern6pt\vrule}\hrule}}

\def\btheta{{\boldsymbol {\theta}}}

\begin{document}
\baselineskip 17pt
\renewcommand {\thepage}{}
\include{titre}
\pagenumbering{arabic}
\title{High-Dimensional Overdispersed Generalized Factor Model with Application to Single-Cell Sequencing Data Analysis}

\begin{spacing}{0.3}%%行间距变为
\author{Jinyu Nie$^1$, Zhilong Qin$^{2}$, Wei Liu$^{3}$\thanks{Corresponding author.   Email: \emph{liuwei8@scu.edu.cn}. The research was partially supported by National
Natural Science Foundation of China (Nos. 11931014).} \\
$^1$Center of Statistical Research and School of Statistics,\\ Southwestern University of Finance and Economics, Chengdu,
 China\\
$^2$Institute of Western China Economic Research, \\ Southwestern University of Finance and Economics, Chengdu, China \\
$^3$School of Mathematics, Sichuan University, Chengdu, China}
\end{spacing}

\date{} 

\maketitle

\begin{abstract}
\baselineskip 18pt
The current high-dimensional linear factor models fail to account for the different types of variables, while high-dimensional nonlinear factor models often overlook the overdispersion present in mixed-type data. However, overdispersion is prevalent in practical applications, particularly in fields like biomedical and genomics studies. To address this practical demand,  we propose an overdispersed generalized factor model (OverGFM) for performing high-dimensional nonlinear factor analysis on overdispersed mixed-type data. Our approach incorporates an additional error term to capture the overdispersion that cannot be accounted for by factors alone. However, this introduces significant computational challenges due to the involvement of two high-dimensional latent random matrices in the nonlinear model. To overcome these challenges, we propose a novel variational EM algorithm that integrates Laplace and Taylor approximations. This algorithm provides iterative explicit solutions for the complex variational parameters and is proven to possess excellent convergence properties. We also develop a criterion based on the singular value ratio to determine the optimal number of factors. Numerical results demonstrate the effectiveness of this criterion. Through comprehensive simulation studies, we show that OverGFM outperforms state-of-the-art methods in terms of estimation accuracy and computational efficiency. Furthermore, we demonstrate the practical merit of our method through its application to two datasets from genomics. To facilitate its usage, we have integrated the implementation of OverGFM into the R package {\it GFM}.

\end{abstract}


\baselineskip 18pt
\section{Introduction}
In recent years, there has been a notable resurgence of high-dimensional factor models, which have proven to be valuable tools for analyzing complex datasets characterized by a large number of variables ~\citep{fan2017sufficient,GFMLiu,jin2021factor}. These models have 
{found} widespread applications across various fields, including economics and finance for asset pricing~\citep{fama1993common}, genomics for cell type identification ~\citep{liu2022joint,liu2023probabilistic}, and social sciences for human ability assessment~\citep{chen2020structured}, among others. The versatility of high-dimensional factor models has positioned them as indispensable tools for addressing the challenges posed by intricate datasets and has paved the way for innovative research and analysis in diverse domains.

High-dimensional factor models provide a powerful framework for capturing the underlying structure and relationships within complex datasets. Through decomposing the observed variables into a reduced number of latent factors, these models enable dimension reduction and facilitate the extraction of meaningful information. The latent factors effectively capture the shared sources of variation across the variables, resulting in a more concise and interpretable representation of the data. In the current  literature, high-dimensional factor models can be divided into two categories: linear factor models and nonlinear factor models.

Bai et al.~\cite{bai2002determining} pioneered the high-dimensional linear factor model (LFM) and significantly advanced the field by establishing estimation theory and demonstrating the consistency of factor number selection. Since then, numerous studies have delved into high-dimensional LFMs~\citep{bai2013principal,fan2017sufficient,li2018embracing, jin2021factor,chen2021quantile}.  LFMs exhibit excellent performance when the relationship between observed variables and factors is linear. However, for high-dimensional data with intricate dependencies, including nonlinearities, LFMs often fall short in terms of goodness of fit~\citep{GFMLiu}.

To address the limitations of high-dimensional LFMs, generalized factor models (GFMs) have been proposed as a class of models that utilize the exponential family of distributions to capture the nonlinear relationship between the high-dimensional observed variables and factors, such as
 Chen et al.~\cite{chen2020structured}, Wang~\cite{wang2020maximum} and Liu et al.~\cite{GFMLiu}. Among these, Chen et al.~\cite{chen2020structured} implicitly assumed a uniform exponential family distribution for all variables, which is not suitable for analyzing mixed-type data. In contrast, Liu et al.~\cite{GFMLiu} and Wang~\cite{wang2020maximum}  considered variable-specific distributions to model mixed-type data, where different variable types corresponded to different distributions. Unfortunately, existing nonlinear factor models are unable to account for overdispersion in mixed-type data, which may result in unsatisfactory {estimation}~\citep{liang1993case,choudhary2022comparison}. Overdispersion is commonly encountered in practice, particularly in biomedical studies involving count responses, where the variability in the observed number of events often exceeds Poisson variability~\citep{liang1993case}. Additionally, overdispersion has been frequently observed in genomics, specifically in the analysis of single-cell RNA sequencing data~\citep{choudhary2022comparison,xia2019spatial}.

% Our model
To overcome the limitations of existing models, we propose  an overdispersed generalized factor model, called OverGFM, which is capable of simultaneously accounting for high-dimensional large-scale mixed data with overdispersion. Building upon the models proposed by Chen et al.~\cite{chen2020structured} and Liu et al.~\cite{GFMLiu}, we formulate a hierarchical structure in OverGFM that incorporates an additional error term to explain overdispersion that cannot be captured by factors alone.
However, OverGFM introduces significant computational challenges stemming from multiple factors. Firstly, it incorporates two high-dimensional latent random matrices, which contribute substantially to the computational complexity. Moreover, the model's inherent nonlinearity adds an additional layer of complexity. To address these challenges, we introduce a variational EM (VEM) algorithm for implementing our model. The VEM algorithm combines Laplace and Taylor approximations, providing iterative explicit solutions for the complex variational parameters. Notably, our proposed VEM algorithm exhibits a high computational efficiency with linear complexity concerning sample size and variable dimension. 
We have theoretically proved the convergence of the proposed VEM algorithm. Furthermore, we develop a criterion based on the singular value ratio to determine the number of factors. 
In simulation studies, OverGFM showed improved estimation accuracy and remarkable computational efficiency in comparison to existing methods. Finally, we employed OverGFM to analyze two sets of single-cell sequencing data. The results unequivocally showcase its capacity  in delivering invaluable biological insights within the genomics field, alongside its impressive computational scalability when addressing vast and intricate datasets.

The remaining sections of the paper are organized as follows. In Section \ref{sec:model}, we provide an introduction to the model setup of OverGFM. Next, in Section \ref{sec:est}, we present the estimation method, specifically focusing on the variational EM algorithm of OverGFM, as well as the procedure for selecting tuning parameter. To evaluate the performance of OverGFM, we conduct simulation studies in Section \ref{sec:simu} and analyze real data in Section \ref{sec:real}. In Section \ref{sec:dis}, we briefly discuss potential avenues for further research in this field. Technical proofs and additional numerical results are provided in the Supplementary Materials. Furthermore, we have seamlessly integrated OverGFM into an efficient and user-friendly R package, conveniently accessible at \url{https://github.com/feiyoung/GFM}.

\section{Model setup}\label{sec:model}
Suppose that the observations $\{\x_i, i = 1, \cdots, n\}$, are independent and identically distributed (i.i.d.), where $\x_i=(x_{i1}, \cdots, x_{ip})^{\trans}$ are variables of mixed types including continuous, binary, count variables, etc. Without loss of generality, we assume that there are $d$ variable types, and the index set of variables for each type $s$ is denoted by $G_s, s= 1, \cdots, d$.
We consider an overdispersed generalized factor model given by a hierarchical formulation,
\begin{eqnarray}
\label{eq:xymodel}
&& x_{ij}|y_{ij} \sim EF(g_s(y_{ij})),   \\
&& y_{ij} = a_{i} +  \b_j^{\trans}\f_i+\mu_j + \varepsilon_{ij}, \  j \in G_s, \ 1 \leq i \leq n,\label{eq:ymodel}
\end{eqnarray}
where {$EF(\cdot)$} is an exponential family distribution and $g_s()$ is called mean function for variable type $s$. For example, if $x_{ij}$ is a continuous variable, then $EF(g_s(y_{ij}))= N(y_{ij}, 0)$, a degenerated normal distribution, i.e., $x_{ij}=y_{ij}$, and $g_s(y)=y$; if $x_{ij}$ is a count variable, then $EF(g_s(y_{ij}))=Poisson(\exp(y_{ij}))$ and $g_s(y)=\exp(y)$; and if $x_{ij}$ is a binary variable, then $EF(g_s(y_{ij}))=Bernoulli(\frac{1}{1+\exp(-y_{ij})})$ and $g_s(y)=\frac{1}{1+\exp(-y)}$. $a_{i}$ is a known offset term for unit $i$, $\f_i \in \mathbb{R}^{q}$ is a vector called latent factors, and $\b_j$ is the corresponding loading vector and $\mu_j$ is an intercept. 
The most significant difference between OverGFM and existing GFMs~\citep{chen2020structured, GFMLiu} is  that OverGFM can account for the extra variations in $x_{ij}$ not explained by factors. This is done with $\varepsilon_{ij}\stackrel{i.i.d.}\sim N(0, \lambda_j)$, which considers these extra variations called  overdispersion~\citep{liang1993case,choudhary2022comparison}. Numerical findings demonstrate that this model design gives OverGFM a performance edge over existing GFMs.

Similar to Liu et al.~\cite{GFMLiu}, we mainly consider three variable types: continuous, count and binomial variables since they are popular in practice and the estimation procedure and the corresponding algorithm can be established  similarly for other types belonging to the exponential family. Without loss of generality, let us assume types 1--3 corresponds to continuous,  count and binomial variables, and denote $p_s=|G_s|$, where $|\cdot|$ is the cardinality of a set.  The models of \eqref{eq:xymodel}--\eqref{eq:ymodel} for these variable types are explicitly written as
\begin{eqnarray}
% \nonumber to remove numbering (before each equation)
  && x_{ij} = y_{ij}, j\in G_1, \quad\quad x_{ij}|y_{ij} \sim Poisson(\exp(y_{ij})) ,  j\in G_2,  \nonumber \\
  && P(x_{ij}=k|y_{ij}) = C_{n_j}^k p_{ij}^k (1-p_{ij})^{n_j-k},p_{ij}=\frac{1}{1+\exp(-y_{ij})}, j\in G_3, \label{eq:G3x} \\
  && y_{ij} =  a_{i} +  \b_j^{\trans}\f_i+\mu_j +\varepsilon_{ij},  j\in G_1 \cup G_2 \cup G_3, \label{eq:ymodelspecific}
\end{eqnarray}
where $n_j$ is the number of trials for the $j$-th variable such that $j\in G_3$. If $n_j=1$ for all $j$, the binomial variable $x_{ij}$ reduces to the Bernoulli variable with success probability $p_{ij}$.
% In Poisson model, i.e., $j\in G_2$, $a_{i}=0$ or $a_{i}=\ln (\sum_{j\in G_3} x_{ij})$; For $j\in G_1\cup G_3$, $a_{i}=0$. $\h_i \sim N(0, \Sigma)$
Model \eqref{eq:ymodelspecific} is unidentifiable due to the unobservability of $\f_i$~\citep{GFMLiu}. Let $\B= (\b_1, \cdots, \b_p)^{\trans}$ be the loading matrix, $\F=(\f_1, \cdots, \f_n)^{\trans}$ be the latent factor matrix, and $\H=(\h_1, \cdots, \h_n)^{\trans}$ be the realization values of $\F$, i.e., factor score matrix. To make models computationally identifiable, we follow  Bai et al.~\cite{bai2013principal} and Liu et al.~\cite{GFMLiu} to impose two conditions on the factor score matrix and loading matrix: (A1) $\frac{1}{n}\sum_{i=1}^{n} \h_i=\bbo$ and $\frac{1}{n}\H\trans\H = \I_q$, where $\I_q$ is a $q$-by-$q$ identity matrix; and (A2) $\B \trans\B$ is diagonal with decreasing diagonal elements and the first nonzero element in each column of $\B$ is positive.

\section{Estimation}\label{sec:est}
Let $p=\sum_{s=1}^3p_s$, $\X=(x_{ij},i=1,\cdots,n,j=1,\cdots,p)\in \mathbb{R}^{n\times p}$ and $\Y=(y_{ij},i=1,\cdots,n,j\in G_2\cup G_3)\in \mathbb{R}^{n\times (p_2+p_3)}$. Denote ${\btheta} = (\mu_{j},\b_j, \lambda_j, j=1,\cdots, {p}, \h_i, i=1,\cdots, n)$ that is the collection of unknown model parameters. 
The conditional log-likelihood (conditional on the latent factor matrix $\F$) of models \eqref{eq:G3x}--\eqref{eq:ymodelspecific} is derived as
\begin{eqnarray}
% \nonumber to remove numbering (before each equation)
  && l(\btheta;\X,\Y|\F=\H) = \sum_i \sum_{j\in G_1}-\frac{1}{2}\{ (x_{ij} - a_i- \b_j^{\trans}\h_i-\mu_j)^2 /\lambda_j +\ln \lambda_j \} \label{eq:logFullP}\\
  && +\sum_i \sum_{j\in G_2} \left\{ (x_{ij} y_{ij} - \exp(y_{ij}))  -\frac{1}{2}\{ (y_{ij}-a_{i} - \b_j^{\trans}\h_i-\mu_j)^2 /\lambda_j +\ln \lambda_j  \} \right\} \nonumber \\
   && + \sum_i \sum_{j\in G_3} \left\{ (x_{ij}-n_j) y_{ij} - n_j \ln (1+\exp(-y_{ij}))  -\frac{1}{2}\{ (y_{ij}-a_{i} - \b_j^{\trans}\h_i-\mu_j)^2 /\lambda_j +\ln \lambda_j \} \right\}, \nonumber
 % && + \sum_i \sum_{j\in G_3} \left\{ (x_{ij}-n_j) y_{ij} - n_j \ln (1+\exp(-y_{ij}))  -\{ \frac{(y_{ij}- a_i- \b_j^{\trans}\h_i-\mu_j)^2}{2\lambda_j}  + \frac{\ln \lambda_j}{2}  \} \right\}, \nonumber
\end{eqnarray}
by omitting the constant independent of parameters.
There are significant computational challenges associated with the fact that  $\Y$ is a large random matrix and $\H$ is a large unknown matrix. In existing GFMs \citep{chen2020structured, GFMLiu}, only the factor score matrix $\H$ is unobservable. As a result, the computational challenges were addressed by treating the latent factors as "parameters" to maximize the conditional log-likelihood~\citep{chen2020structured, GFMLiu}.
However, this approach is not applicable to our overdispersed GFM because of the additional unobservable large random matrix $\Y$. Therefore, we consider $\Y$  as latent variables handled by expectation-maximization (EM) algorithm, while $\H$ is regarded as a high-dimensional  matrix parameter to be estimated directly.
The EM algorithm \citep{dempster1977maximum} is a powerful and well-developed framework for handling models with latent variables, and involves the posterior distribution of the latent variables in a key step. However, in our model, computing the posterior distribution $P(\Y|\X,\H)$ is extremely challenging due to the high dimensionality of both $\Y$ and $(\X,\H)$, as well as the presence of nonlinear terms for Poisson and binomial variables in the log-likelihood \eqref{eq:logFullP}.
% To unify the forms of updating, let $a_i =(0, a_i, 0)$.

To make the posterior distribution tractable, we utilize a mean field variational family, $q(\Y)$, to approximate $P(\Y|\X, \H)$:
$$q(\Y)= \Pi_{i,j\in G_2\cup G_3} N(y_{ij}; \tau_{ij}, \sigma^2_{ij}).$$
Let ${\bg} = (\tau_{ij},\sigma^2_{ij}, i=1,\cdots, n, j \in G_2\cup G_3)$ that is the collection of unknown variational parameters. In the proposed algorithm,  $\bg$ is solved to seek an optimal approximation in the sense that KL divergence of $q(\Y)$ and $P(\Y|\X, \H)$ is minimized.

Next, we derive the evidence lower bound (ELBO) function, which is given by
\begin{eqnarray*}
% \nonumber to remove numbering (before each equation)
  && ELBO(\btheta;\bg) = \sum_i \sum_{j\in G_1}-\frac{1}{2}\{ [(x_{ij} - \b_j^{\trans}\h_i-\mu_j-a_i )^2] /\lambda_j +\ln \lambda_j \} \\
  && +\sum_i \sum_{j\in G_2} \left.\bigg\{ (x_{ij} \tau_{ij} - \exp(\tau_{ij}+{\frac{\sigma_{ij}^2}{2}}))  -\frac{1}{2}\{ [(\tau_{ij}-a_i  - \b_j^{\trans}\h_i-\mu_j)^2 + \sigma_{ij}^2]/\lambda_j \right.\\\\
  &&  \left. +\ln \lambda_j  \} \right. \bigg\}+ \sum_i \sum_{j\in G_3} \bigg\{ (x_{ij}-n_j) \tau_{ij} - n_j \mathrm{E}_{q(y_{ij})}\ln (1+\exp(-y_{ij}))  \\
   && -\left.  \frac{1}{2}\{ [(\tau_{ij}- \b_j^{\trans}\h_i-\mu_j-a_i )^2 + \sigma_{ij}^2 ]/\lambda_j +\ln \lambda_j  \} \right\} + \frac{1}{2}\sum_{i,j\in G_2 \cup G_3} \ln (\sigma_{ij}^2),
\end{eqnarray*}
where $\mathrm{E}_{q(y_{ij})} F(y_{ij})$ is taking the expectation of $F(y_{ij})$ with respect to the random variable $y_{ij}\sim N(\tau_{ij}, \sigma^2_{ij})$. In the following, we present a variational EM algorithm designed to implement the model.

\subsection{Variational E-step}
Unlike the conventional EM algorithm, the variational EM approach transforms the posterior expectation in the E-step into an optimization problem involving the variational parameters. Then, we introduce how to update the variational parameters $\bg=(\tau_{ij},\sigma^2_{ij},1\leq i\leq n, j \in G_2\cup G_3)$ given model parameters $\btheta$.
However, it is very difficult to evaluate these parameters  because $q(\Y)$ is not a conjugate distribution to $P(\X|\Y)$. We turn to the Laplace approximation~\citep{wang2013variational} to obtain an approximate posterior distribution of $\Y$. Specifically, since $P(\Y|\X) \propto P(\X|\Y) P(\Y|\H)$, a Taylor
approximation around the maximum a posterior point of $P(\X|\Y) P(\Y|\H)$ is adopted to construct a Gaussian proxy for the
posterior.

For $j\in G_2$, $\ln P(\X|\Y) P(\Y|\H) = \sum\limits_i \sum\limits_{j\in G_2}\{x_{ij}y_{ij} - \exp(y_{ij}) - \frac{1}{2}\{(y_{ij}- a_{i}- \mu_j - \b_j^{\trans}\h_i)^2/\lambda_j\}\} +c$, where $c$ is a constant independent of parameters.
Let $f_{ij}(y)=x_{ij}y- \exp(y) - \frac{1}{2} \{(y-a_{i}-\mu_j-\b_j^{\trans}\h_i)^2/\lambda_j\}$, then the posterior mean and variance of $y_{ij}$ can be estimated by
$$\hat \tau_{ij} = \arg\max_{y} f_{ij}(y), \hat \sigma^2_{ij} = -f''_{ij}(\hat\tau_{ij})^{-1}, i=1,\cdots,n,j\in G_2, $$
where $f''_{ij}(y)=-\exp(y) - \lambda_j^{-1}$. The derivation details are provided in Appendix A.1 of Supplementary Materials.

However, maximizing $f_{ij}(y)$ with respect to $y$ is computation-consuming since both $n$ and $p_2$ may be very large. To {improve} the computational efficiency, we further enhance the Laplace approximation by creatively combination with Taylor approximation. Specifically, before maximizing  $f_{ij}(y)$,  we apply the Taylor approximation to the exponential term of $f_{ij}(y)$. Recalling $g_2(y) =  \exp(y)$ and by Taylor's theorem, we can approximate $g_2(y)$ by
expanding around $y_0$, i.e.,
$g_2(y) \approx \tilde g_2(y) = g_2(y_0) + g_2'(y_0) (y-y_0) + \frac{1}{2} g_2''(y_0) (y-y_0)^2.$
Substituting $\tilde g_2(y)$ into $f_{ij}(y)$ and taking derivative to $y$, we obtain an explicit iterative value of $\hat\tau_{ij}$ as well as $\hat\sigma_{ij}^2$,
\begin{equation}\label{eq:updateY}
 \hat\tau_{ij}=\frac{x_{ij} - \exp(y_0) (1-y_0)+\lambda_j^{-1}\tilde z_{ij}}{\lambda_j^{-1} + \exp(y_0)},    \hat\sigma_{ij}^2=\frac{1}{\lambda_j^{-1}+\exp(\hat\tau_{ij})},
\end{equation}
where $y_0$ is taken as the previous iterative value of $\tau_{ij}$,  and $\tilde z_{ij} = a_{i}+ \mu_j + \b_j^{\trans}\h_i$.

Similarly, for $j \in G_3$,  let $f_{ij}(y)=(x_{ij}-n_j)y - n_j\ln (1+\exp(-y)) - \frac{1}{2}\{(y-\b_j^T\h_i - \mu_j-a_i)^2/\lambda_j\}$, then
$$\hat \tau_{ij} = \arg\max_{y} f_{ij}(y), \hat \sigma^2_{ij} = -f''_{ij}(\hat\tau_{ij})^{-1}. $$
The explicit form of  $f''_{ij}(y)=-n_j g_3(y)(1-g_3(y)) - \lambda_j^{-1}$ with $g_3(y)=\frac{1}{1+\exp(-y)}$. Let $h(y)= \ln g_3(y)$, then the second-order Taylor expansion is $h(y)\approx \tilde h(y)=h (y_0) + h'(y_0) (y-y_0) + \frac{1}{2} h''(y_0) (y-y_0)^2$, where $h'(y)=(1-g_3(y))$ and $h''(y)=-g_3(y) (1-g_3(y))$. Substituting $\tilde h(y)$ into $f_{ij}(y)$ and taking derivative to $y$, we obtain
\begin{equation}\label{eq:updateY3}
\hat \tau_{ij} = \frac{x_{ij}-n_jg_3(y_0)+n_j y_0 g_3(y_0) (1-g_3(y_0)) + \lambda_j^{-1}\tilde z_{ij}}{\lambda_j^{-1} + n_j g_3(y_0)(1- g_3(y_0))}, \hat \sigma^2_{ij} = \frac{1}{\lambda_j^{-1}+ n_jg_3(\hat \tau_{ij})(1-g_3(\hat \tau_{ij}))  }.
\end{equation}

The iterative closed-form solutions, as denoted in equations \eqref{eq:updateY} and \eqref{eq:updateY3}, play a pivotal role in achieving computational efficiency. Until now, the variational E-step is finished, then the variational M-step is considered to update the model parameters $\btheta$ by fixing the variational parameter $\bg$.

\subsection{Variational M-step}

Taking derivative of $ELBO(\btheta;\bg)$  with respect to each model parameter and setting it to zero, we obtain the updated formula:
\begin{eqnarray}
% \nonumber to remove numbering (before each equation)
   \b_j &=& (\H^{\trans}\H)^{-1} \sum_i \h_i(\bar x_{ij}- \mu_j), \label{eq:updata_bj} \\
   \h_i &=& (\B^{\trans}\Lambda^{-1}\B)^{-1} \sum_j \b_j(\bar x_{ij}- \mu_{j})/\lambda_j, \\
   \mu_j &=& \frac{1}{n} \sum_i (\bar x_{ij} - \b_j^T\h_i), \\
   \lambda_j &=& \frac{1}{n} \sum_{i} \{(\bar x_{ij} - \b_j^T\h_i-\mu_j)^2 + \sigma^2_{ij}\}. \label{eq:updata_lambdaj}
\end{eqnarray}
where $\bar x_{ij}=x_{ij}- a_i$ if $j\in G_1$, $\bar x_{ij}=\tau_{ij}-a_i$ if $j\in G_2\cup G_3$, $\sigma_{ij}^2= 0$ for $j\in G_1$, and $\Lambda=\mathrm{diag}(\lambda_1,\cdots, 
{\lambda_p})$. According to Equations \eqref{eq:updateY}--\eqref{eq:updata_lambdaj}, it is  straightforward  to implement the variational EM algorithm summarized in Algorithm \ref{alg:vem}. The implementation details are given in Appendix A.3 of Supplementary Materials.

The proposed variational EM algorithm aims to iteratively maximize the evidence lower bound function by optimizing block coordinate directions. The parameter space $\mathcal{G}$ is defined as the set of parameters satisfying Conditions (A1) to (A2). In the Supplementary Materials, a formal proof is provided to demonstrate the convergence of the iterative algorithm. The result is stated as follows:
\begin{Theorem}
If conditions {(B1)–(B2)} in the Supplementary Materials hold,  given the proposed variational EM algorithm, we have that
all the limit points of $(\btheta^{(t)}, \bg^{(t)})$ are local maxima of $ELBO(\btheta, \bg)$ in the parameter space $\mathcal{G}$, and $ELBO(\btheta, \bg)$ converges monotonically to $L^{*}=ELBO(\btheta^{*}, \bg^{*})$ for some $(\btheta^{*}, \bg^{*})\in\mathcal{G}^{*}$, where $\mathcal{G}^{*}=\{\mbox{set of local maxima in the interior}$ $\mbox{ of }~\mathcal{G}\}.$
\end{Theorem}

\begin{algorithm}
	\renewcommand{\algorithmicrequire}{\textbf{Input:}}
	\renewcommand{\algorithmicensure}{\textbf{Output:}}
	\caption{The proposed variational EM algorithm for OverGFM}
	\label{alg:vem}
	\begin{algorithmic}[1]
		\REQUIRE $\X$, $q$, maximum iterations $maxIter$, relative tolerance of $ELBO$  ($epsELBO$).
		\ENSURE  $\wh\H, \wh\B, \wh\bmu, \wh \Sigma, \wh\Lambda$
		\STATE Initialize $\bg^{(0)} = (\tau^{(0)}_{ij},\sigma^{2,(0)}_{ij},i \leq n, j \leq  p)$ and $\btheta^{(0)}=(\B^{(0)},\bmu^{(0)},\H^{(0)},\Lambda^{(0)}) $.
		\FOR{ each $t = 1, \cdots, maxIter $}
		\STATE Update variational parameters $\bg^{(t)}$ based on Equations~\eqref{eq:updateY}--\eqref{eq:updateY3};
		\STATE Update model parameters $\btheta^{(t)}$ based on Equations~\eqref{eq:updata_bj}--\eqref{eq:updata_lambdaj};
        \STATE Evaluate the evidence lower bound $ELBO_t=ELBO(\btheta^{(t)}, \bg^{(t)})$.
        \IF {$|ELBO_t - ELBO_{t-1}|/ |ELBO_{t-1}| < epsELBO$}
        \STATE  break;
        \ENDIF
		\ENDFOR
		\STATE  Exert the identifiability conditions (A1)--(A2) on $\H^{(t)}$ and $\B^{(t)}$.
		\STATE \textbf{return} $\wh\H=\H^{(t)}, \wh\B=\B^{(t)}, \wh\bmu=\bmu^{(t)}, \wh\Lambda=\Lambda^{(t)}$.
	\end{algorithmic}
\end{algorithm}

\subsection{Selection of the number of factors}\label{sec:selectq}
The number of factors ($q$) is an undetermined tuning parameter that requires selection. To tackle this issue, we present a simple and effective method based on singular value ratio (SVR) that can be easily implemented.

Our proposed SVR method draws inspiration from the eigenvalue ratio-based approach commonly employed to determine the number of factors in linear factor models \citep{ahn2013eigenvalue}. In this method, the estimation of $q$ is carried out using $\hat q = \arg\max_{k\leq q_{max}} \frac{\kappa_k(\hat\Phi_x)}{\kappa_{k+1}(\hat\Phi_x)}$, where $\hat\Phi_x$ represents the sample covariance of $\x_i$ within the linear factor model framework, and $\kappa_k(\hat\Phi_x)$ denotes the $k$-th largest eigenvalue of $\hat\Phi_x$.
The underlying concept behind this approach can be intuitively understood as follows. Assuming that the true number of factors is $q$, the eigenvalues $\kappa_{k}(\hat\Phi_{x})$ for $k>q$ primarily originate from the error term's variance, $\varepsilon_i$. Consequently, $\kappa_{k}(\hat\Phi_{x})$ for $k>q$ is noticeably smaller compared to $\kappa_{q}(\hat\Phi_{x})$, resulting in a considerably large value for $\frac{\kappa_{q}(\hat\Phi_{x})}{\kappa_{q+1}(\hat\Phi_{x})}$. However, applying this approach directly to our nonlinear factor model becomes challenging due to the absence of a linear structure between the observed variables and the factors.

\begin{figure}[H]
  \centering
  \includegraphics[width=14cm]{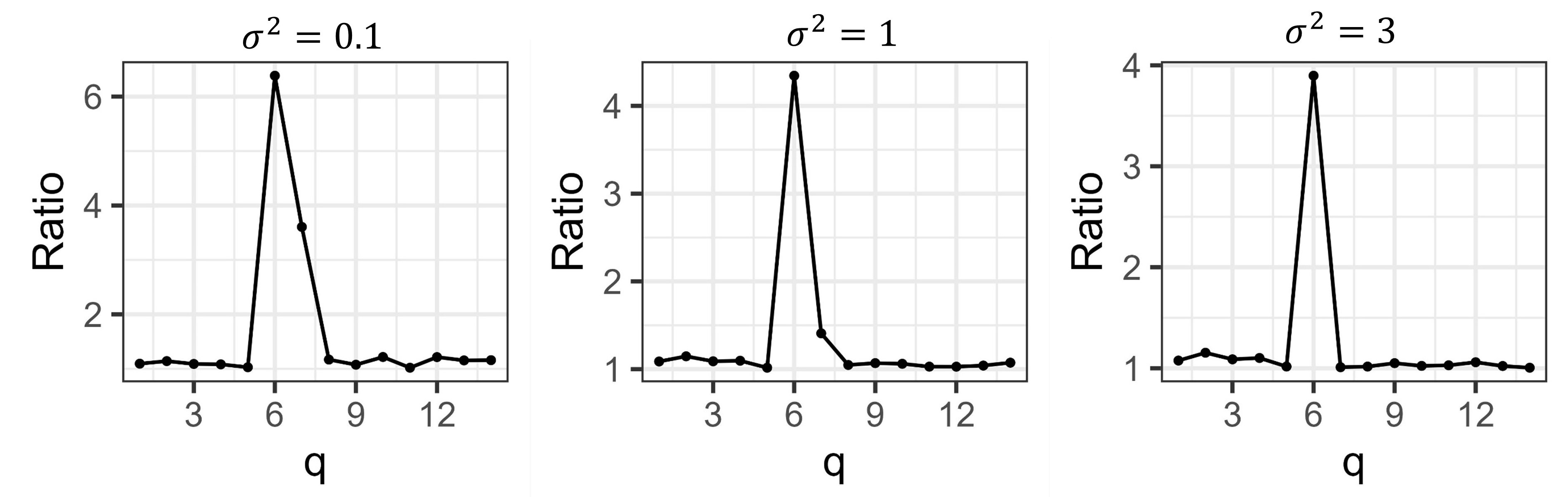}
  \caption{The  singular value ratio of $\widehat{\mathbf{B}}$ obtained by OverGFM from a  random sample under three different settings of the error variance, where $\lambda_j=\sigma^2 \in \{0.1, 1, 3\}, (n,p)=(300, 300)$ and the true value $q=6$.}\label{fig:q}
\end{figure}

Similar to Chen et al.~\cite{chen2021nonlinear}, we introduce a surrogate, denoted by $\hat\Phi_{hb}$, for $\hat\Phi_x$,  to tackle this issue. It is defined as the sample covariance matrix of $\wh\B\wh\h_i$.  Due to the identifiable conditions satisfied by $\wh\H$ and $\wh\B$, we have $\hat\Phi_{hb}=\wh\B\wh\B^{\trans}$. Let $q_{\max}$ be the upper bound for $q$. First, we fit our model using $q=q_{\max}$, then define the estimator of $q$ as
$\hat q = \arg\max_{k\leq q_{max}} \frac{\nu_k(\wh\B)}{\nu_{k+1}(\wh\B)}$, where $\nu_k(\wh\B)$ is the $k$-largest singular value of $\wh\B$.  This method is referred to as the singular value ratio (SVR) based method. The empirical results depicted in Figure \ref{fig:q}, obtained from Scenario 4 of Section 4, demonstrate the performance of the SVR method and its potential to  identify the true value of $q$. As error's variance increases, the maximum singular value ratio decreases, indicating an increase in the difficulty of accurately identifying the true value of $q$. More comprehensive investigation is conducted in Section 4.

\section{Simulation study}\label{sec:simu}
%In this section, we conduct extensive simulation studies to examine the performance of the proposed method.

In this section, we showcase the effectiveness of the proposed OverGFM through simulation studies involving 200 realizations. We compare OverGFM with various state-of-the-art methods from the current literature. They include

\begin{itemize}
    \item[(1)] Generalized factor model~\citep{GFMLiu} implemented in the R package {\bf GFM};
    \item[(2)] Multi-response reduced-rank  regression model ~\citep[MRRR,][]{luo2018leveraging} implemented in {\bf rrpack} R package;
    \item[(3)] Principal component analysis for data with mix of qualitative and quantitative variables~\citep{chavent2022multivariate}, implemented in the R package {\bf PCAmixdata};
    \item[(4)] Generalized PCA (GPCA)~\citep{landgraf2020generalized} implemented in the R package {\bf generalizedPCA};
    \item[(5)] High-dimensional LFM~\citep{bai2002determining} implemented in the R package {\bf GFM};
    \item[(6)] Poisson PCA~\citep{kenney2021poisson} implemented in the R package {\bf PoissonPCA};
    \item[(7)] PLNPCA~\citep{chiquet2018variational} implemented in the R package {\bf PLNmodels};
\end{itemize}
% ~(8) Logistic PCA~\citep{landgraf2020dimensionality} implemented in the R package {\bf logisticPCA};
Among the aforementioned methods, methods (1)-(3) are capable of handling mixed-type data. Method (4) can only analyze single-type data, including continuous, count, and categorical types. Method (5) is specifically designed for analyzing continuous variables, widely recognized as a benchmark with broad applications, notably in economics~\citep{bai2021matrix,fan2017sufficient} and genomics~\citep{argelaguet2020mofa+,liu2022joint}, whereas methods (6) and (7) excel at analyzing count data. 

We  evaluate OverGFM in a total of eight scenarios. In scenarios 1-3, our main focus is comparing OverGFM with methods (1)-(3) and LFM, as LFM is widely utilized in practical applications~\citep{jin2021factor,liu2023probabilistic}. We generate data with a mixed-type of three variable types for these scenarios. In scenarios 4 and 5, we investigate the performance of the proposed SVR method in  selecting the number of factors and the estimation performance under misselected $q$, respectively.  In scenario 6, we investigate the computational efficiency of OverGFM by comparing it with other methods. In scenario 7, we focus on special cases where data is generated using a combination of two variable types or a single variable type, aiming to compare OverGFM with methods (4), (6) and (7). In scenario 8, we delve into the interconnection between OverGFM and GFM. To conserve space, the results (Table S1--S3, Figure S2) pertaining to scenarios 7--8 are deferred to Appendix B of the Supplementary Materials. In implementing the compared methods, we maintain the default settings and solely adjust the argument for the number of factors/principal components (PCs). To facilitate a fair comparison, we set the number of factors/PCs to the true value for all methods.

In scenarios 1--3, we generate data from models \eqref{eq:xymodel} and \eqref{eq:ymodel}, i.e., $x_{ij}|y_{ij} \sim EF(g_s(y_{ij}))$,  and $y_{ij} = a_{i} +  \b_j^{\trans}\h_i+\mu_j + \varepsilon_{ij}$, and consider the  mix of three different variable types: continuous, count and  binary variables, i.e., $g_1(y)=y, g_2(y) = \exp(y)$ and $g_3(y)=1/(1+\exp(-y))$. Without loss of generality, we set the offset $a_i=0$ for all $i$'s. We set the number of variables of these three variable types to $\lfloor\frac{p}{3}\rfloor, \lfloor\frac{p}{3}\rfloor$ and $p-2\lfloor\frac{p}{3}\rfloor$, respectively. Next, we generate $\breve{\B}=(\breve{b}_{jk})\in \mathbb{R}^{p\times q}$ with $\breve{b}_{jk}\stackrel{i.i.d.}\sim N(0,1)$. Let $\breve{\B}_s$  be the submatrix of loading for variable type $s$. We generate $\bar{\B}_s=\rho_s \breve{\B}_s$, then construct $\bar{\B}=(\bar{\B}_1\trans,\bar{\B}_2\trans, \bar{\B}_3\trans)\trans$, where $\rho_s$ controls the signal strength of each variable type. To obtain the singular value decomposition (SVD), we decompose  $\bar{\B}=U_2 \Lambda_2 V_2^{\trans}$. Next, we define $\B_0= U_2 \Lambda_2$. We then generate $\breve{\h}_i$ from $N(\0_q,(0.5^{|i-j|})_{q\times q})$ and denote $\breve{\H}=(\breve{\h}_1,\cdots,\breve{\h}_n)^{\trans}$, perform column orthogonality for $\breve{\H}$ to obtain $\bar\H$, and set $\H_0=\bar\H V_2^{\trans}/\sqrt{n} $ such that $\H_0^{\trans}\H_0/n=\I_q$.  Note that $\H_0$ and $\B_0$  satisfy the identifiable conditions (A1)--(A2) given in Section \ref{sec:model}. Finally, We  generate $\mu_j=0.4z_j, z_j \sim N(0,1)$, and $\varepsilon_{ij} \sim N(0,\sigma^2)$.   In all scenarios, we fix the number of latent factors ($q$) to 6.

We assess the estimation accuracy of the intercept-loading matrix $\Upsilon=(\bmu, \B)\in \mathbb{R}^{p\times (q+1)}$ and the factor matrix $\H$ by utilizing the commonly-used trace statistic~\citep{doz2012quasi} that measures the distance of the column space spanned by two matrices. For the factor score matrix,  the trace statistic, denoted as $\mathrm{Tr}(\wh\H,\H_0)$, is defined as
$\mathrm{Tr}(\wh\H,\H_0)=\frac{\mathrm{Tr}(\H_0\wh\H^{\trans} (\wh\H^{\trans}\wh\H)^{-1} \wh\H^{\trans}\H_0)}{\mathrm{Tr}(\H_0^{\trans}\H_0)}$.
The trace statistic yields a value between 0 and 1, with a higher value indicating greater accuracy in estimation.

\begin{figure}[H]
  \centering
  \includegraphics[width=14cm]{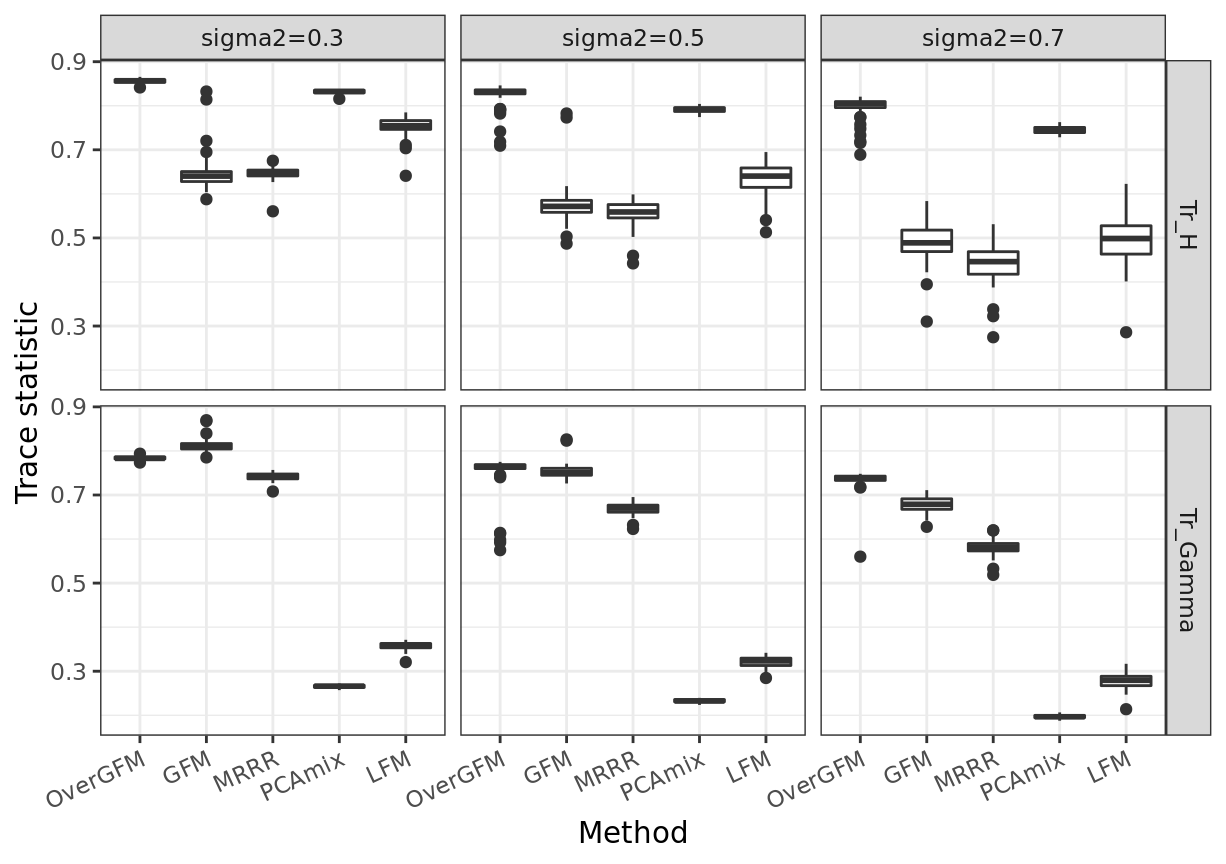}
  \caption{Comparison of estimation accuracy among OverGFM and other methods under  secenario 1 with three mixed-type variables, where $(n,p) = (500,500), q = 6, \sigma^2 \in \{0.3,0.5,0.7\}$,  Tr\_H and Tr\_Gamma denote the trace statistics with respect to $\H$ and $\Upsilon$, respectively. }\label{fig:secenario1}
\end{figure}

\noindent\underline{Scenario 1}.
%comparison under different noise
First, we aim to explore the impact of overdispersion on the performance of OverGFM and other methods under consideration. In this scenario, we set $(n,p)=(500,500)$ and $(\rho_1,\rho_2,\rho_3) = (0.05,0.2,0.1)$ as fixed values, while varying the overdispersion parameter $\sigma^2$ within the grid $\{0.3,0.5,0.7\}$.  We compare  OverGFM with other methods such as GFM, MRRR, PCAmix and LFM since only GFM, MRRR and PCAmix are able to handle the mixed-type data while LFM is widely used in practice despite not explicitly considering variable types. As shown in Figure \ref{fig:secenario1}, the results clearly demonstrate that OverGFM outperforms the other methods under consideration. This superiority becomes even more evident as the values of $\sigma^2$ increase. Notably, PCAmix shows good accuracy in estimating the factor matrix, but it performs poorly when it comes to estimating the loading-intercept matrix.  This limitation can be attributed to the fact that PCAmix simply combines PCA and multiple correspondence analysis (MCA) to handle mixed-type data, and does not distinguish between count and continuous variables~\citep{chavent2022multivariate}. Furthermore, we observe that LFM performs poorly in  estimating the intercept-loading matrix since LFM focuses solely on modeling linear dependencies among variables at the mean scale. This underscores the significance of capturing nonlinear dependencies between mixed-type variables.Additionally, we find that GFM, which does not account for overdispersion, exhibits inadequate performance
in factor estimation. This emphasizes the significance of incorporating overdispersion into the modeling approach to achieve accurate results. 

In addition, we investigate the performance of the proposed method in comparison to its competitors when the overdispersion mechanism is incorrectly specified and there are highly heavy tails present. The results (Figure S1) show that OverGFM is not only flexible to the overdispersion but also robust to the heavy-tail data and model misspecification, making it
a highly attractive and favorable choice in practical applications; see Appendix B.1 in Supplementary Materials.

\begin{figure}
\centering
\subfigure[]{
\includegraphics[width=12cm]{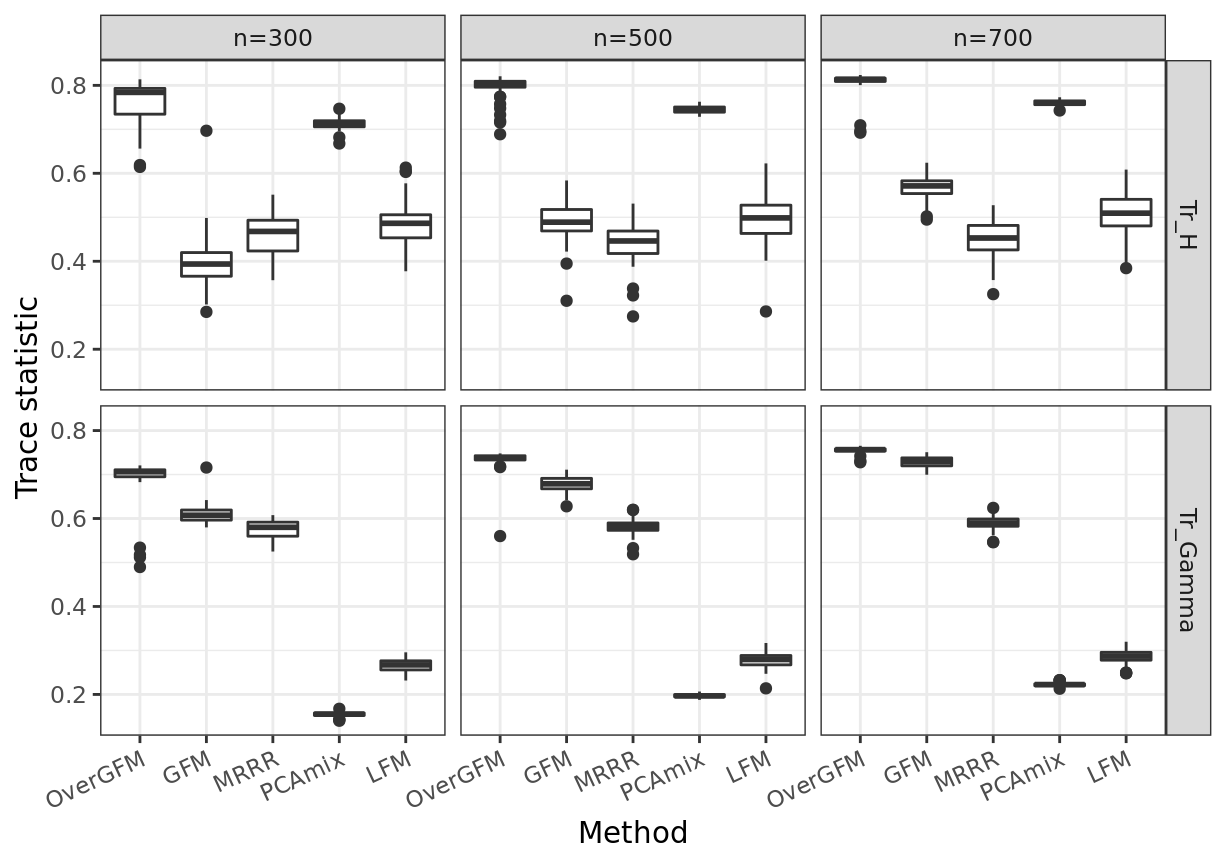}
\label{fig:difn}
}
\subfigure[]{
\includegraphics[width=12cm]{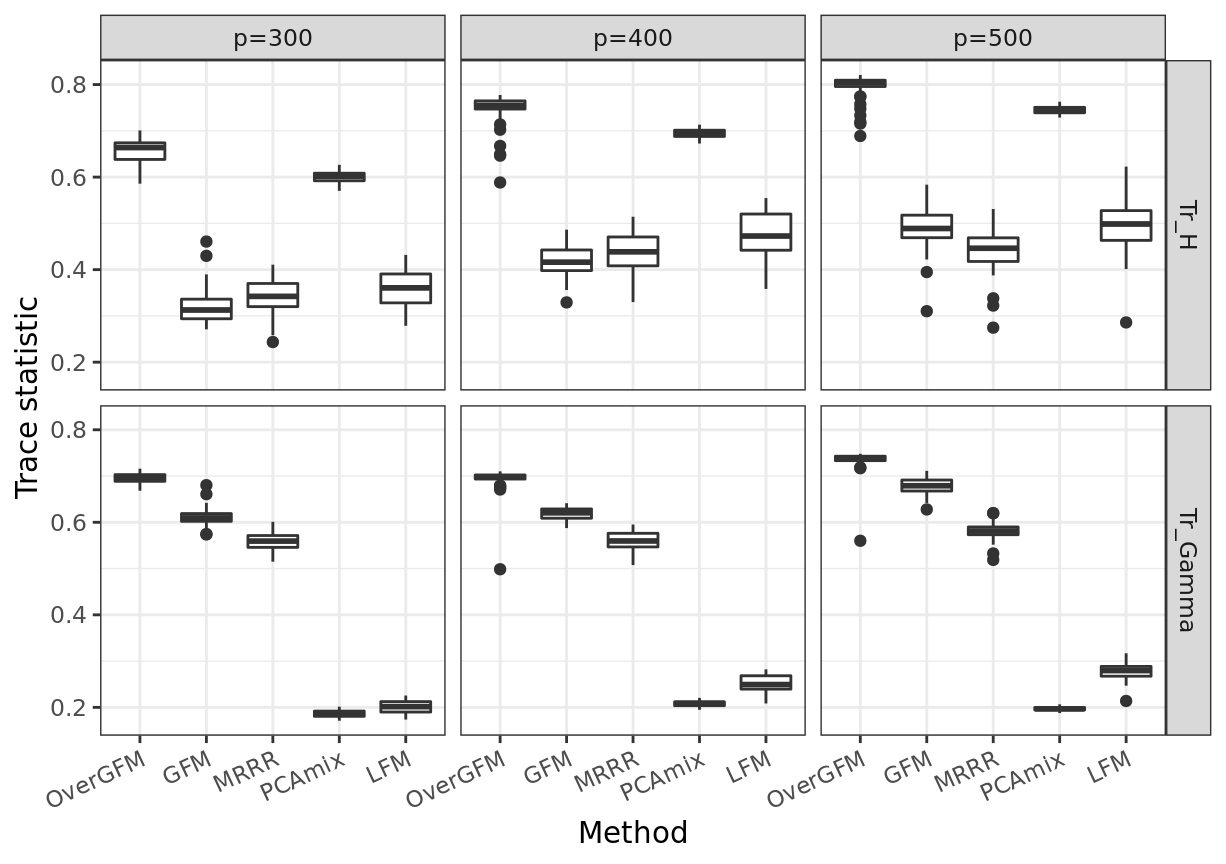}
\label{fig:difp}
}
\caption{(a) \& (b): Comparison of  estimation accuracy among  OverGFM and other methods under  scenario 2 with varying sample size, i.e.,  $p = 500, n = (300,500,700)$,  and varying variable dimension, i.e.,$n = 500, p = (300,400,500)$.}\label{fig:secnario2Fig}
\end{figure}

\noindent\underline{Scenario 2}. To assess the estimation accuracy as the sample size ($n$) or the number of variables ($p$) increases, we generate data with a fixed number of variables ($p=500$) and varying sample sizes ($n \in \{300, 500, 700\}$), or a fixed sample size ($n=500$) and varying numbers of variables ($p \in \{300, 400, 500\}$). We set the overdispersion parameter to $\sigma^2 = 0.7$ and other setting same as that in the scenario 1. We compare  OverGFM with  GFM, MRRR, PCAmix and LFM. Figure \ref{fig:secnario2Fig} demonstrates that OverGFM surpasses other methods in terms of estimation accuracy for factor and loading-intercept matrices across various structural dimensions.  As $n$ or $p$ increases, the performance of OverGFM becomes better. We observe that the enhancement in intercept-loading matrix estimation is more sensitive to the sample size $n$, while the improvement in factor matrix estimation is more sensitive to the variable dimension $p$. This distinction arises because each intercept-loading vector is estimated based on information from $n$ individuals, whereas each factor vector is estimated using information from $p$ variables. 

\begin{figure}[H]
  \centering
  \includegraphics[width=14cm]{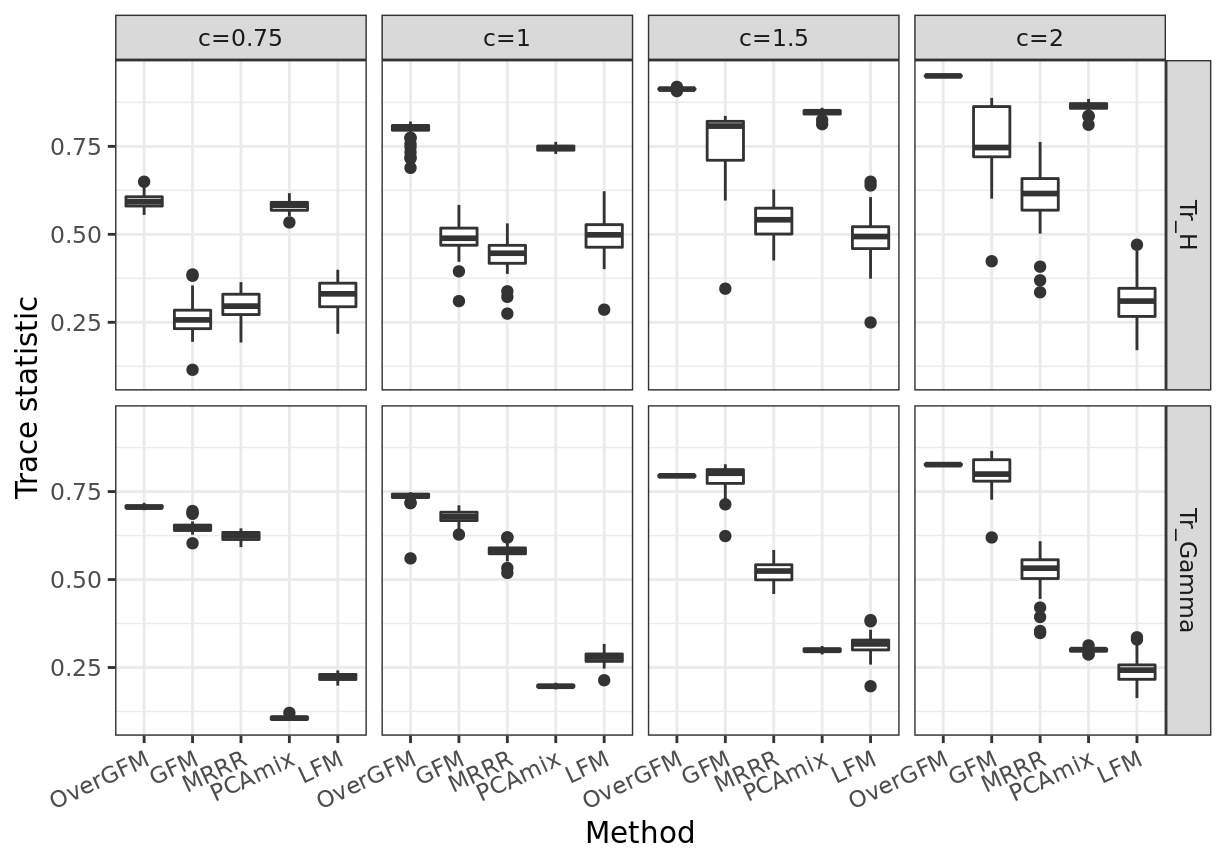}
  \caption{Comparison of  estimation accuracy among  OverGFM and other methods under  secenario 3  with three mixed-type variables, where $(n,p) = (500,500), q=6$, and $(\rho_1, \rho_2, \rho_3)=c \times (0.05,0.2,0.1)$ with $c \in \{0.75, 1, 1.5, 2 \}$.}\label{fig:scenario3}
\end{figure}

\noindent\underline{Scenario 3}.
We then investigate the influence of the signal strength in loading matrix on the performance of OverGFM. Specifically, we fix $\sigma^2 = 0.7$ and $(n,p)=(500,500)$ while increase the signal strength by setting $(\rho_1, \rho_2, \rho_3) = c\times(0.05, 0.2, 0.1)$ with $c\in \{0.75, 1, 1.5, 2\}$. Figure \ref{fig:scenario3} clearly illustrates that as the signal strength increases, both OverGFM and other methods that account for variable types (GFM and PCAmix) {exhibit an upward trend in estimation peformance while LFM not.} Importantly, OverGFM consistently outperforms the other methods under comparison.

\begin{figure}
  \centering
  \includegraphics[width=14cm]{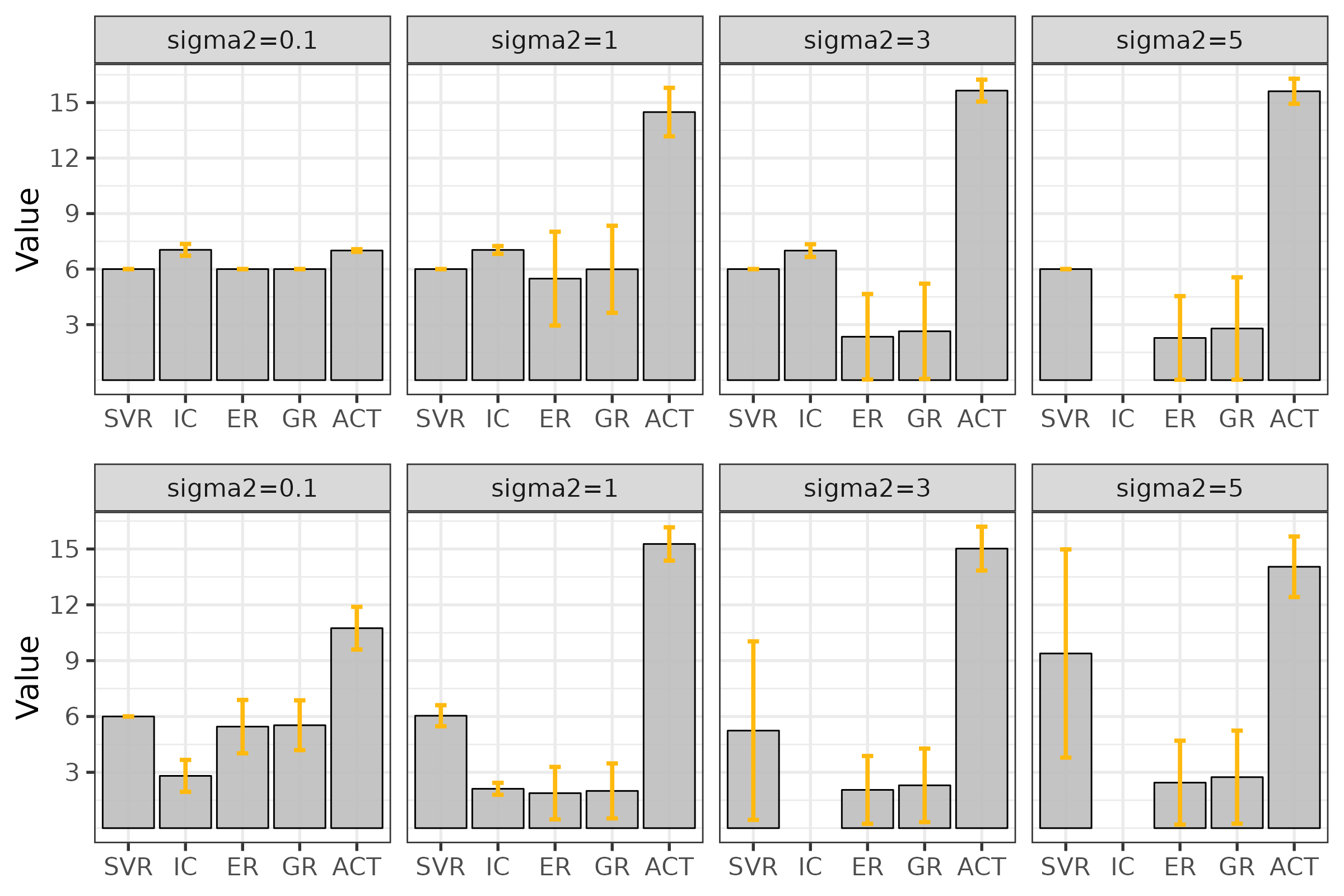}
  \caption{Comparison of  factor number identification performance between the proposed SVR method and four alternative methods under $(n,p)=(300,300)$ and $q=6$. Upper panel: mix of three variable types: normal, Poisson and binary. Bottom panel: mix of two variable types: Poisson and binary. The algorithm for the IC method breaks down when $\sigma^2\geq 3$.}\label{fig:selectq}
\end{figure}

\noindent\underline{Scenario 4}. Furthermore, we investigate the performance of the SVR criterion given in Section \ref{sec:selectq} that selects the number of factors.  We compare our proposed SVR method with  the information criterion (IC) for GFM in \cite{GFMLiu}, eigenvalue ratio (ER) and ratio of the growth rates (GR) based methods in  \cite{ahn2013eigenvalue}, and  adjusted correlation thresholding (ACT) method in \cite{fan2022estimating}. Note that the ER, GR, and ACT methods were proposed specifically for the linear factor model framework. To make comparison fair, we set the same range $\{1,2,\cdots, 15\}$ as candidated values for SVR and other compared methods. For our analysis, we examine two cases for data generation. {In case 1}, we generate data incorporating a combination of three variable types, following the same way as described in Scenario 1. In case 2, we generate data containing a mix of Poisson and binary variables, using the same data generating process as outlined in scenario 7. Case 1 has stronger signal than case 2 in terms of the variable type since case 1 includes the continuous variables. For both cases, we keep the dimensions fixed at $(n,p)=(300, 300)$ while varying the error variance ($\sigma^2$) across the grid values of $\{0.1, 1, 3,5\}$ to investigate the impact of the overdispersion. Figure  \ref{fig:selectq} provides valuable insights. {In case 1} where there is a strong signal in the variable type, when the  overdispersion is low ($\sigma^2=0.1$), all methods successfully identify the underlying structure dimension. Nevertheless, when overdispersion reaches $3$, only SVR and IC prove effective, whereas ACT, ER, and GR falter in capturing the true structure. Further amplifying the overdispersion to $5$, SVR stands alone in its effectiveness. Moving on to case 2, characterized by a weak signal in the variable type, SVR, ER, and GR perform well when the overdispersion is low ($\sigma^2=0.1$). However, with the escalation of overdispersion, only SVR retains the capacity to identify the true number of factors. However, a subsequent increase in overdispersion renders all methods ineffective, attributable to insufficient signals. Notably, the algorithm for the IC method breaks down when the overdispersion is high ($\sigma^2\geq 3$).

\begin{figure}[H]
  \centering
  \includegraphics[width=14cm]{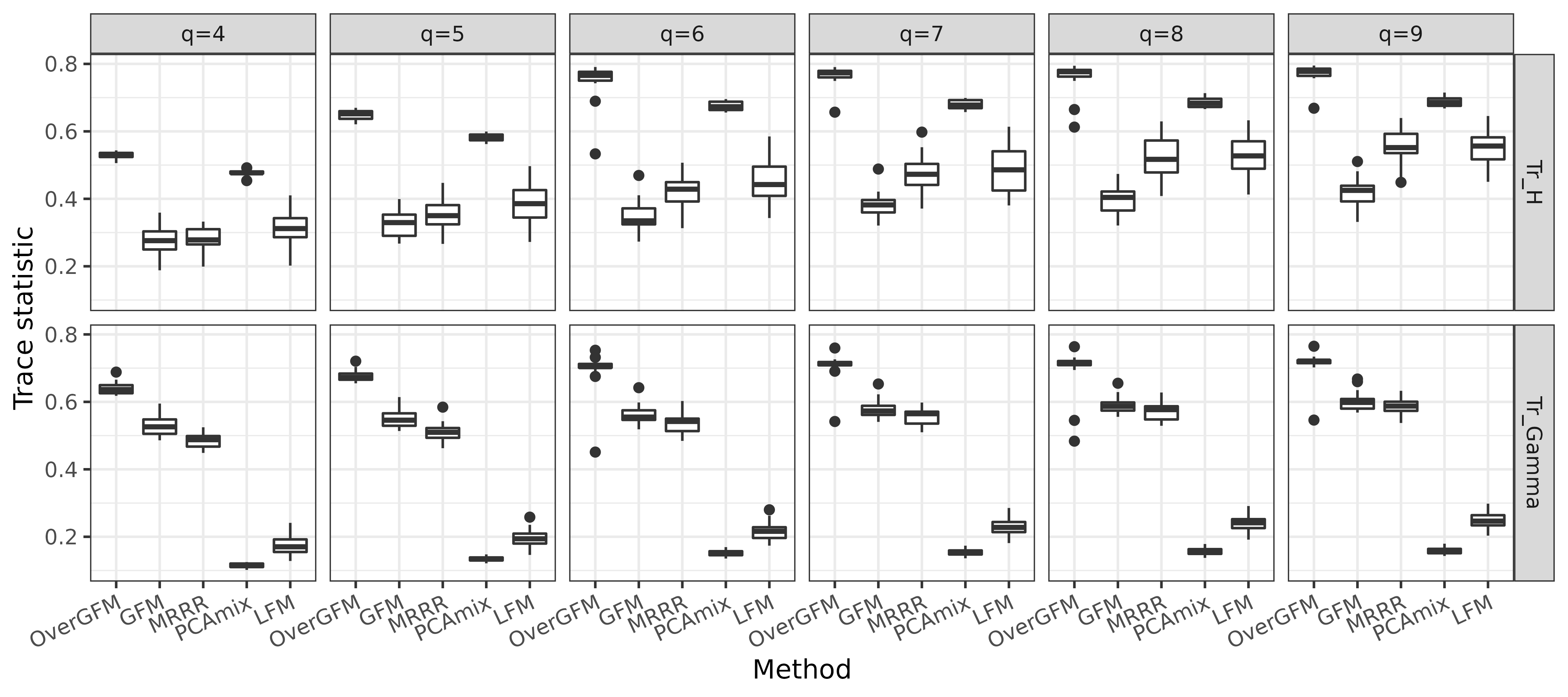}
  \caption{Comparison of  estimation performance for OverGFM and four other methods when the number of factors is misselected from  $\{4,5,6,7,8,9\}$ and  the true factor number is $q = 6$. }\label{fig:secenario7}
\end{figure}

% : comparison different latent factor number
\noindent\underline{Scenario 5}. 
Inadequate data signal may lead to incorrect estimation of the number of factors, as showed in Scenario 4. We designed this scenario to study how estimators perform when the number of factors is misselected. We set $(n,p) = (300,500)$ and $q=6$, and vary the selected $q$ from $\{4,5,6,7,8,9\}$. Other settings remained the same as in Scenario 1. In
Figure \ref{fig:secenario7}, we observe that
the estimation accuracy of OverGFM is consistently higher than that of  the compared methods across {all selected $q$s}. More importantly, we find that the estimation accuracy of OverGFM with over-selected  $q$ significantly outperforms  that with under-selected $q$. These results offer valuable guidance for practitioners using OverGFM. Based on the SVR  method, users can opt for larger values of  $q$ to achieve more reliable and robust applications of the model.

\begin{figure}
  \centering
  \includegraphics[width=14cm]{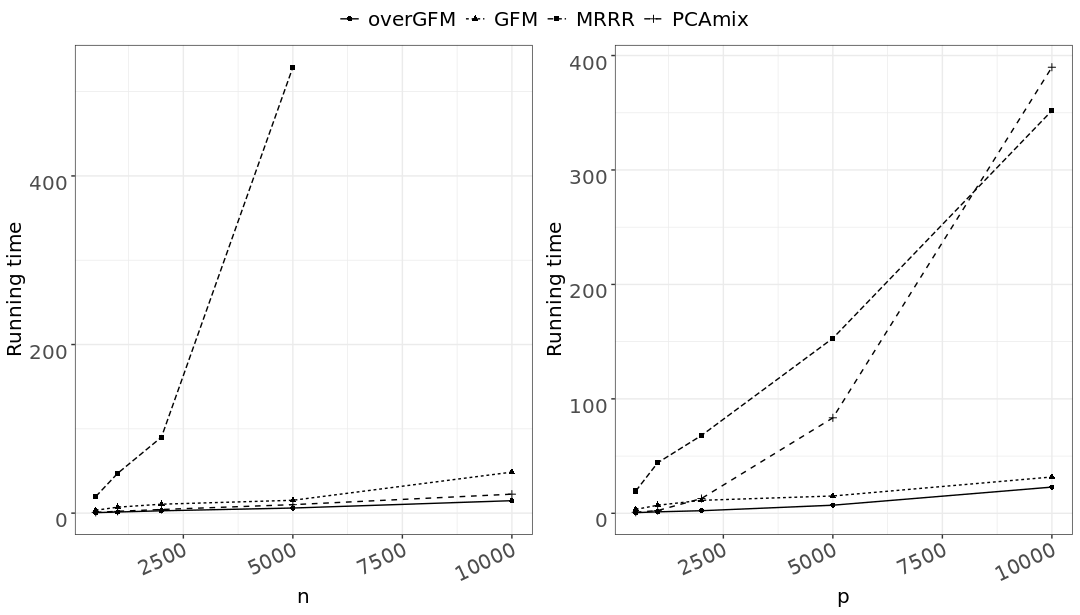}
  %\hspace{1in}
  %\includegraphics[width=8cm]{secenario5-2.png}
  \caption{Comparison of  the average running time over 20 runs for OverGFM and three other  methods: GFM, MRRR and PCAmix. Left panel:  $p=500, n \in \{500, 1000,  2000, 5000,10000\}$. Right panel: $n=500, p \in \{500, 1000,  2000, 5000, 10000\}$.}\label{fig:secenario5-12}
\end{figure}

% : comparison computational efficiency
\noindent\underline{Scenario 6}. Finally, we assess the computational efficiency of OverGFM in comparison to  GFM, MRRR and PCAmix since only these methods account for mixed-type variables. We consider two cases by generating data as the same as scenario 2. In case 1, we fix $p$ at $500$ while vary  $n$ from $500$ to $10000$; in case 2, we fix
$n$ at $500$ while vary  $p$ from $500$ to $10000$ (Figure \ref{fig:secenario5-12}).
Figure \ref{fig:secenario5-12} displays the average running time over {20 runs} for each method. Remarkably, OverGFM exhibits linear computational complexity with respect to both $n$ and $p$, and outperforms the other methods, particularly MRRR and PCAmix. Our observations reveal that MRRR exhibits poor scalability concerning both the sample size and variable dimension. As these parameters increase, MRRR's running time experiences a substantial surge. Specifically, when the sample size reaches 10000, MRRR takes approximately 3000 seconds which are too long to display. This finding highlights the limited scalability of MRRR.  Additionally, our analysis suggests that PCAmix struggles to handle increasing variable dimensions efficiently. In this scenario, our results clearly demonstrate that OverGFM outperforms other methods in terms of computational efficiency, making it a highly attractive and favorable choice.

\section{Real data analysis}\label{sec:real}
In this section, we showcase the successful application of OverGFM in analyzing single-cell sequencing data within the genomics field. This includes the utilization of OverGFM on both a single-cell RNA sequencing (scRNA-seq) dataset and a single-cell multimodal sequencing dataset. The results demonstrate the effectiveness and versatility of OverGFM in handling diverse genomics data types.
\subsection{scRNA-seq data of mouse olfactory bulb}
In the analysis of single-cell RNA sequencing (scRNA-seq) data, the common presence of overdispersion is noticeable across various studies~\citep{kharchenko2021triumphs,hafemeister2019normalization,choudhary2022comparison}. To show the utility  of OverGFM, we apply it to analyze the scRNA-seq data obtained from the mouse olfactory bulb (OB), a neural structure involved in processing olfactory information and enabling the sense of smell.

Investigating the cell type heterogeneity and identifying marker genes within the OB holds substantial significance in this context. The data set, which can be accessed at \url{https://panglaodb.se/view_data.php?sra=SRA667466&srs=SRS3060025}, consists of 1,578 cells and 24,109 genes measured using the 10X chromium technology. The expression levels of each gene are represented as count reads, and the website provides cell cluster labels for all cells. This enables us to assess the performance of OverGFM and compare it with other methods in terms of feature extraction, by examining the association between the extracted features and the annotated cell clusters.  Following the guideline of scRNA-seq data analysis~\citep{stuart2019comprehensive}, we first select the top 1,000 highly variable genes of high quality. By computing the variance-to-mean ratio, a widely utilized metric for assessing overdispersion, for each gene, we noted a pronounced overdispersion. Please refer to Supplementary Figure S3, which aligns with  the observations in the previous studies. Based on the 1,000 count variables, we construct the continuous and binary variables to form a data with three variable types. To obtain continuous variables, the specific log-normalization was performed on a gene expression read count value $z_{ij}$, i.e., $x_{ij}=\ln (1+z_{ij})$ that avoids the the issue of $z_{ij}=0$. To create binary variable, we assign a value of $x_{ij}=1$ if $z_{ij}>0$, and $x_{ij}=0$ if $z_{ij}=0$. %We perform log-normalization to transform the 1,000 count variables into continuous variables. Additionally, we generate 1,000 binary variables by assigning a value of 1 if the read count for a gene {exceeds} 0, and 0 otherwise. 
Our primary
objective is to investigate the dimension reduction performance of OverGFM by comparison with three other methods that can handle the mixed-type data, i.e., GFM, MRRR, PCAmix. Furthermore, we also compared OverGFM to LFM, which is commonly used in practice.

\begin{figure}[H]
\centering\includegraphics[width=\textwidth]{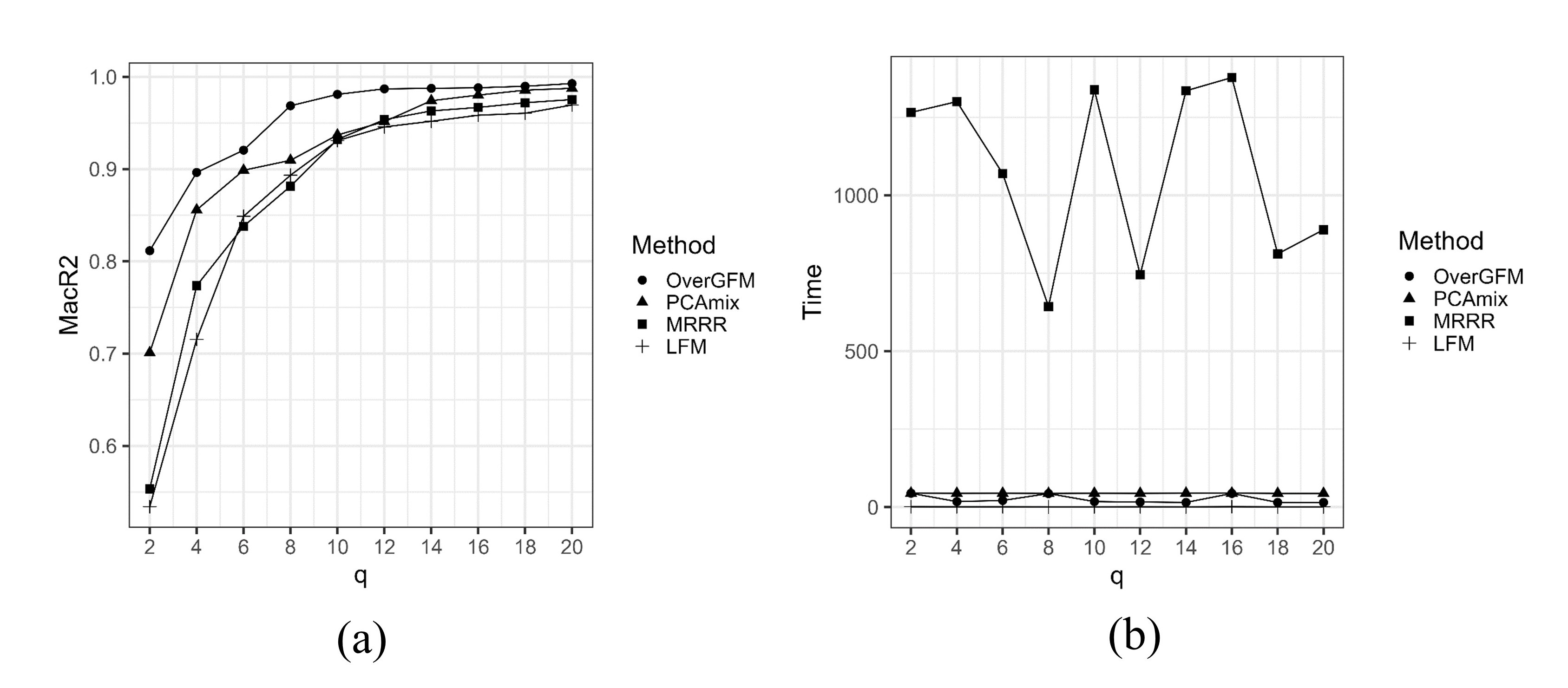}
  \caption{Comparison of   OverGFM and three other  methods: MRRR, PCAmix and LFM. (a) The adjusted McFadden's pseudo R$^2$ across different numbers of factors; (b) The running time
 across different numbers of factors.}\label{fig:macR2etc}
\end{figure}

% compare different methods
To evaluate the performance of OverGFM and other methods, we fit each method with different numbers of factors by varying $q \in \{2, 4, \cdots, 18, 20\}$. Subsequently, we calculate the adjusted McFadden's pseudo R$^2$~\citep{mcfadden1987regression}  between the extracted features and the annotated cell clusters for each fitted model. This metric provides a measure of the amount of biological information captured by the features, where a higher value indicates superior performance in dimension reduction.
Notably, we observe that GFM encountered difficulties when applied to this data. Similar to simulations, its algorithm displayed instability and ultimately failed. Therefore, in Figure \ref{fig:macR2etc}(a), we only present the McFadden's pseudo R$^2$ results for OverGFM, MRRR, PCAmix, and LFM. Remarkably, these findings indicate that OverGFM outperformed the other methods across the range of $q$ values considered.
Furthermore, we recorded the running time for each method in Figure \ref{fig:macR2etc}(b). The results consistently demonstrated that OverGFM exhibited the highest efficiency compared to MRRR and PCAmix. This finding aligns with the conclusions drawn from our simulations. By varying the selection of highly variable genes from 1,000 to 5,000, we confirm that OverGFM demonstrates remarkable scalability with respect to the variable dimension, completing computations in less than 160 seconds even for $p=15,000$, as illustrated in Supplementary Figure S4. Additionally, the robustness of OverGFM is verified by selecting 2,000 highly variable genes and comparing it with other methods, as shown in Supplementary Figure S5.

\begin{figure}
\centering\includegraphics[width=\textwidth]{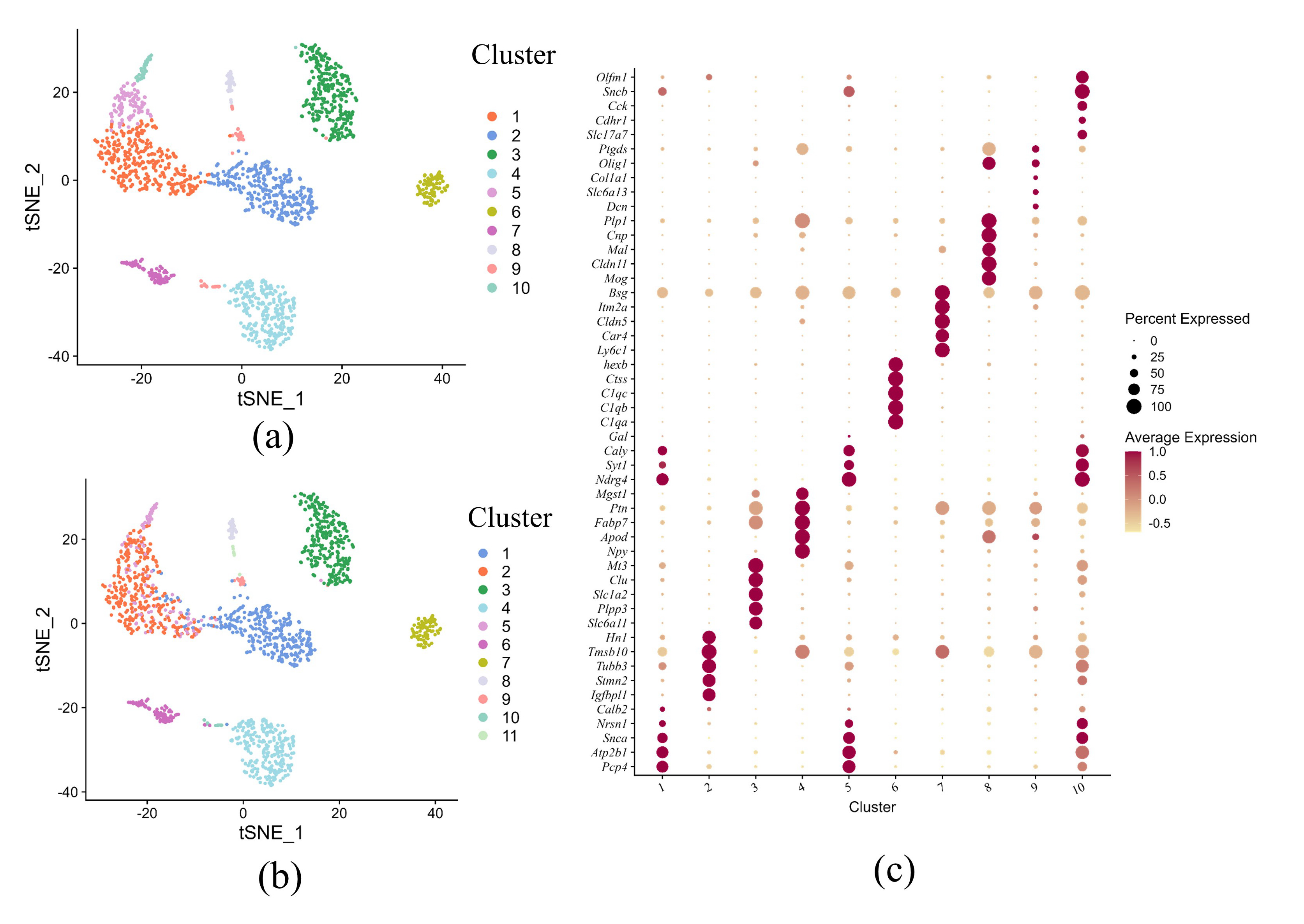}
  \caption{Downstream analysis using the extracted features from OverGFM. (a) tSNE plot for visualizing the cluster labels of Louvain clustering using $\wh\H$; (b) tSNE plot for visualizing the annotated cluster labels; (c) Dot plot of top five marker genes for the 13 identified clusters by
 Louvain clustering. Y-axis:
 marker genes.}\label{fig:clusteretc}
\end{figure}

% Downstream analysis
Next, we show the valuable utility of the estimated factor matrix obtained from OverGFM in essential downstream analyses, such as cell type identification and differential gene expression analysis. Based on the proposed SVR criterion, we select the number of factors $\hat q=6$ by setting $q_{\max}=15$. Then we fit OverGFM to obtain $6$-dimensional features, denoted by $\wh\H$. We perform the Louvain clustering on $\wh\H$, which is widely used in
single-cell RNA sequencing data analysis~\citep{stuart2019comprehensive}, and identify 10 distinct
cell clusters.  By visualizing the identified clusters on two-dimensional tSNE embeddings~\citep{van2008visualizing} extracted from $\wh\H$, we can observe a clear separation of distinct cell clusters (Figure \ref{fig:clusteretc}(a)). Moreover, upon comparing the identified clusters with the annotated cell clusters, we observe a close alignment between them (Figure \ref{fig:clusteretc}(b)). In addition, we identify  two subtypes of the annotated cluster 2 in Figure \ref{fig:clusteretc}(b), which are clusters 1 and 5 in Figure \ref{fig:clusteretc}(a). Based on the identified clusters, we detect the differentially expressed genes for each cluster. The dot plot (Figure \ref{fig:clusteretc}(c)) demonstrates the clear separation of the identified top marker genes across the 10 distinct clusters, providing further evidence of the high-quality cell clustering achieved using $\wh\H$.
Using the marker genes and the cell-type database, PanglaoDB (https://panglaodb.se/), we manually map the 10 identified cell clusters to specific cell types. Table \ref{tab:celltype} provides the cell types for the identified cell clusters, along with the marker genes that determine these cell types. Our manual annotations reveal that clusters 1 and 5 represent subtypes of Purkinje neurons. To explore the roles of these two subtypes in cellular functional mechanisms, we performed Gene Ontology
gene set enrichment analysis to investigate the functions associated with each subtype; see Figure S6 and Appendix B.2 in Supplementary Materials.  All these results conclusively demonstrate that the estimated factor matrix of OverGFM is highly valuable and beneficial for scRNA-seq data analysis.

\begin{table}[!ht]
\renewcommand\tabcolsep{1.0pt} % 调整表格列间的宽度
    \centering\scriptsize
    \caption{Identification of cell types for the ten cell clusters based on the detected marker genes.}\label{tab:celltype}
    \begin{tabular}{lccccc}
    \hline
        Cell cluster & 1 & 2 & 3 & 4 & 5  \\ \hline
        
        Cell type & Purkinje neurons & Interneurons & Astrocytes & Schwann cells &  Purkinje neurons \\ \hline
        Marker genes & \makecell{Pcp4,\\ Atp2b1,\\ Snca} & \makecell{Stmn2,\\ Tmsb10,\\ Tubb3} &  \makecell{Mt3,\\Clu, \\ Plpp3,\\ Slc1a2}  & \makecell{Apod,\\ Fabp7,\\ Ptn} &  \makecell{Pcp4,\\ Atp2b1,\\ Snca} \\ \hline
        Cell cluster & 6 & 7 & 8 & 9 & 10 ~ \\ \hline
        Cell type & Microglia &  Endothelial cells& Oligodendrocytes & \makecell{Oligodendrocyte\\ progenitor cells} & Interneurons \\ \hline
        Marker genes & \makecell{C1qa, \\ C1qb,\\ C1qc}&  \makecell{ Bsg,\\ Car4,\\ Ly6c1}  &\makecell{Cldn11, \\ Cnp, \\ Mal}  & \makecell{ Olig1, \\ Ptgds} & \makecell{Cck,\\  Slc17a7,\\ Sncb}  \\ \hline
    \end{tabular}
\end{table}
%% identify the functional difference of the subtypes?

\subsection{SNARE-seq data of mouse cerebral cortex}
\begin{figure}
\centering\includegraphics[width=\textwidth]{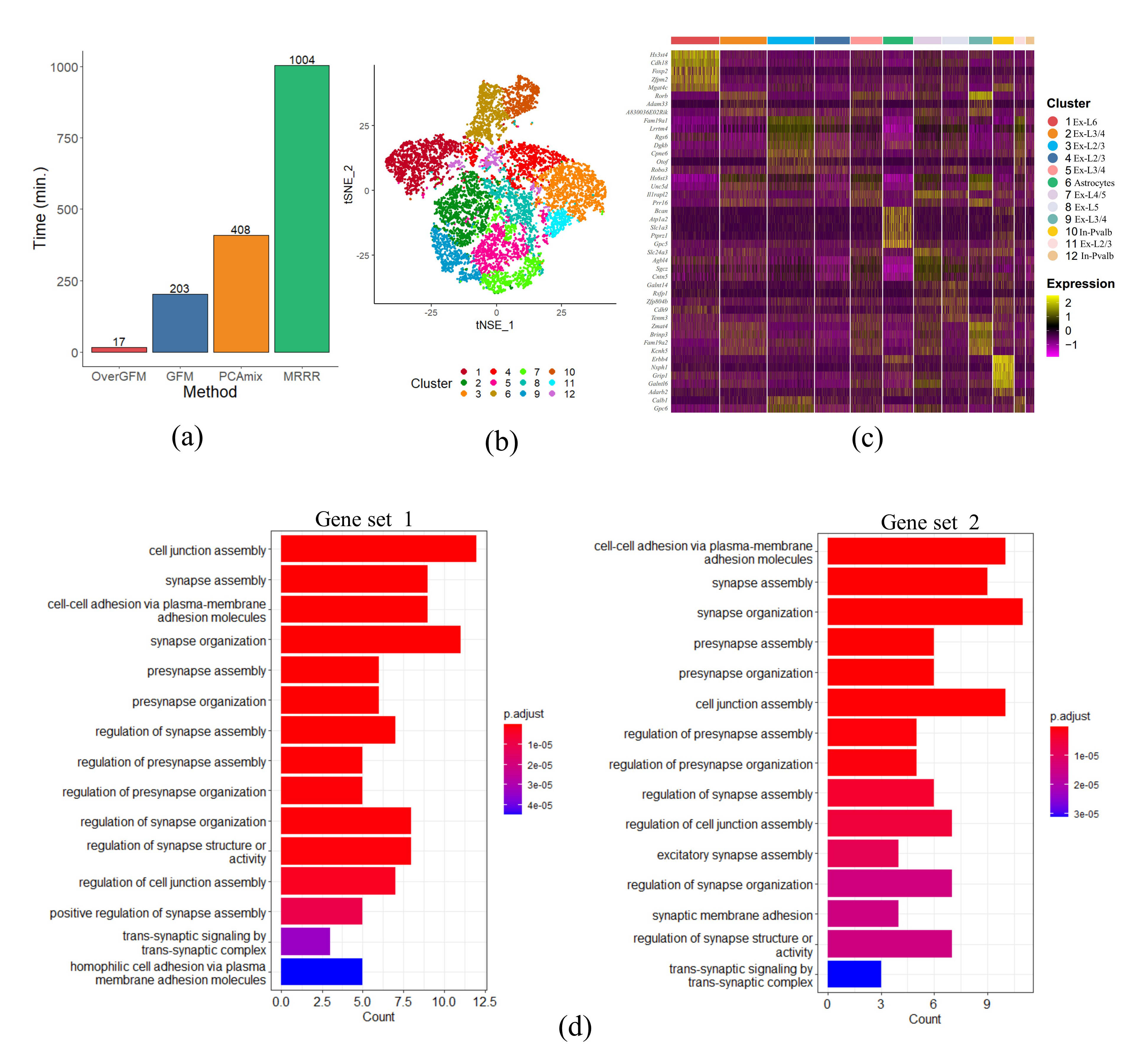}
  \caption{SNARE-seq data analysis using OverGFM. (a) Comparison of running time for OverGFM and other methods that can handle mixed-type data; (b) tSNE plot for visualizing the  cluster labels based on the estimated factors by OverGFM; (c) Heatmap of expresson levels for top five marker genes of the 12 clusters. The clusters are annotated by the marker genes. Ex: excitatory neurons, In: Inhibitory neurons, and L6: layer 6; (d) Barplots showing significant pathways from gene set enrichment analysis for two gene sets in the biological process category of the GO database.}\label{fig:snareseq_main}
\end{figure}

We integrate multi-modal data using OverGFM, measured by SNARE-seq technology~\citep{chen2019high} in adult mouse cerebral cortex, which is available by GEO accession number GSE126074. The dataset has 10309 cells with two modalities: chromatin accessibility (binary) and mRNA expression (read counts). The chromatin accessibility has 244544 sites, and mRNA expression has 33160 genes~\citep{chen2019high}. To streamline the analysis, we first filter out sites with fewer than 500 nonzero entries across all cells, resulting in a selection of 10085 sites. Additionally, we identify the top 2000 highly variable genes, bringing the total features for subsequent analysis to 12085. As no ground truth about the cell clusters is available, our focus lies in comparing the computational cost of OverGFM with other methods capable of handling mixed-type data. Utilizing the SVR criterion, we select the number of factors as $\hat q=5$ by setting $q_{\max}=15$. For fair comparison, we fix the number of factors for the other compared methods at 5 as well. Figure \ref{fig:snareseq_main}(a) illustrates the notable computational advantage of OverGFM, requiring only 17 minutes, while GFM, PCAmix, and MRRR demand 203, 408, and 1004 minutes, respectively.

Based on the $\wh\H$ obtained through OverGFM, we employ Louvain clustering to group cells exhibiting similar chromatin accessibility and gene expression profiles into 12 distinct clusters. To visualize these clusters effectively, we employ two-dimensional tSNEs. Figure \ref{fig:snareseq_main}(b) showcases the impressive separation achieved for different clusters.
Subsequently, we perform gene differential expression analysis to pinpoint marker genes for each cluster, which can serve as identifying characteristics. This analysis is depicted in Figure \ref{fig:snareseq_main}(c).
To determine the cell types represented by these clusters, we compare our identified marker genes with those documented in the published paper~\citep{chen2019high}. The results are presented in Figure \ref{fig:snareseq_main}(c), revealing the cell types associated with each cluster.

Except for cell typing, 
in this data analysis, we demonstrate that the estimated loading matrix $\wh\B\in \mathrm{R}^{p \times 5}$ by OverGFM enhances the discovery of important gene sets, where $p=2000+10085$. 
Specifically, we identify these crucial gene sets by ranking the magnitudes of loadings for each of the five directions. This approach is motivated by the fact that the magnitudes of loadings reflect the contribution of genes to the feature extraction process. Genes with larger loadings are deemed to contain more critical and informative characteristics, making them essential for the analysis and interpretation of the data.
For each column of the $2000$-by-$5$ submatrix in the upper block of $\wh\B$, we select the top 50 genes with the largest loading magnitudes to form a gene set; see Table S4 and Figure S7(a) in Supplementary Materials. Next, we conduct gene set enrichment analysis for the biological process category in the Gene Ontology database to explore the functions of these gene sets. Interestingly, Figure \ref{fig:snareseq_main}(d) and Supplementary Figure S7(b) show that gene sets 1-4, corresponding to loadings 1-4, are significantly enriched in biological processes related to cell junction assembly, cell-cell adhesion via plasma-membrane adhesion molecules, and synapse organization. These findings suggest that these gene sets play a vital role in establishing and maintaining the complex organization of the highly specialized brain region.

This data example unequivocally demonstrates that OverGFM is an immensely valuable and advantageous tool for multi-modal sequencing data analysis.

\section{Discussion}\label{sec:dis}
We have introduced a novel statistical model named OverGFM, designed for the analysis of high-dimensional overdispersed mixed-type data. This model proves particularly valuable when dealing with scenarios where both the sample size (n) and variable dimension (p) are substantial. To address the computational challenges of large-scale data, we have developed a computationally efficient variational EM (VEM) algorithm. The VEM algorithm exhibits linear computational complexity with respect to sample size and variable dimension, ensuring its scalability. It offers straightforward implementation with explicit iterative solutions for all parameters and guarantees convergence, supported by theoretical proofs. Moreover, we  developed a singular value
ratio based method to determine the number of factors. In a series of numerical experiments, we have demonstrated that OverGFM surpasses existing methods, achieving superior estimation accuracy while reducing computation time. This makes OverGFM a compelling choice for analyzing extensive mixed-type datasets.
Additionally, our application of OverGFM in the analysis of scRNA-seq and SNARE-seq data has proved its efficacy in unveiling the underlying structure of complex genomics data. We are confident that OverGFM holds the potential to enable essential discoveries not only in the field of genomics but also across other scientific domains. Its versatility and efficiency make it a promising tool for data analysis in various research areas.

A straightforward extension for OverGFM involves managing the high-dimensional mixed-type data with additional high-dimensional covariates.  This extension would enable the model to explore the association between mixed-type variables and the extra covariates while also considering the presence of unobserved latent factors that cannot be explained by the covariates alone. We plan to pursue this direction in our future work, as it holds the promise of further enhancing the model's capabilities and broadening its applicability to a wider range of real-world data analysis scenarios.

\bibliographystyle{WileyNJD-AMA}%{plain}%

\bibliography{reflib1}

\end{document}

\title{Supplementary Materials to ``High-Dimensional Overdispersed Generalized Factor Model with Application to Single-Cell Sequencing Data Analysis"}
\date{}
\author{Jinyu Nie$^1$, Zhilong Qin$^{2}$, Wei Liu$^{3*}$, \\
$^1$Center of Statistical Research and School of Statistics,\\ Southwestern University of Finance and Economics, Chengdu,
 China\\
$^2$Institute of Western China Economic Research, \\ Southwestern University of Finance and Economics, Chengdu, China \\
$^3$School of Mathematics,  Sichuan University, Chengdu, China}
\maketitle

\baselineskip 18pt
\setcounter{section}{0}
\Appendix
\section{Materials about the algorithm}

\subsection{Details of the variational E-step}
To enhance comprehension of our algorithm, we present the rationale behind estimating the posterior mean and variance in the variational E-step.

By Bayes's formula, we known the posterior distribution of $P(\Y|\X) \propto P(\X|\Y) P(\Y|\H)=\Pi_{j\in G_1} (\Pi_i^n P(x_{ij}|\h_i)) \times \Pi_{j\in {G_2 \cup G_3}} (\Pi_i^n P(x_{ij}|y_{ij}) P(y_{ij}|\h_i))$. Thus, when $j\in G_2\cup G_3$, the true posterior density of $y_{ij}$ is $q_{0,ij}(y_{ij})\equiv P(x_{ij}|y_{ij}) P(y_{ij}|\h_i) /c_{ij}$, where $c_{ij}$ is a normalization constant that ensures $q_{0,ij}(y_{ij})$  being a density function. If we are based on this true posterior, the computation is intractable in the E-step. Thus, we use a Gaussian density $N(\tau_{ij}, \sigma^2_{ij})$ to approximate $q_{0,ij}(y_{ij})$, which is equivalent to a Taylor approximation around the maximum a posterior point of $q_{0,ij}(y_{ij})$. Thus, the maximum a posterior point of $q_{0,ij}(y_{ij})$ is the approximated posterior mean, i.e., $\tau_{ij}=\arg\max_{y}q_{0,ij}(y)=\arg\max_{y} f_{ij}(y)$ and the approximated variance  is the inverse of the second derivative of $-f_{ij}(y)$ evaluated at the maximum a posterior point $\tau_{ij}$, which is just the Laplace approximation.

\subsection{Proofs of Theorem 3.1}
Let $F(\btheta, \bg)$ be the mapping function of the proposed variational EM algorithm, which means that the algorithm
generates the sequences $\{\btheta^{(t)}, \bg^{(t)}\}$ by $(\btheta^{(t+1)}, \bg^{(t+1)})=F(\btheta^{(t)}, \bg^{(t)})$. Let
$\mathcal{G}_0=\{(\btheta, \bg) \in \mathcal{G}: ELBO(\btheta, \bg)\geq ELBO(\btheta^{(0)}, \bg^{(0)})\}$. Then we present a definition of a closed mapping.
\begin{Def}(\cite{Luenberger2016}, page 199)
A point-to-set mapping $G$ from $\mathcal{X}$ to $\mathcal{Y}$ is said to be closed at $x\in \mathcal{X}$ if the assumptions 1)  $\lim_{k\rightarrow \infty} x_k \rightarrow x$, 2) $ \lim_{k\rightarrow \infty} y_k \rightarrow y$ and 3) $y_k \in G(x_k)$ imply $y\in G(x)$. Moreover, $G$ is said to be closed over $\mathcal{X}$ if $G$ is closed at every point of $\mathcal{X}$.
\end{Def}

Let $A$ and $B$ be two sets, and we define $A\backslash B$ as the set difference between sets $A$ and $B$.
Then we pose two conditions:
\begin{description}
  \item[(B1)] $\mathcal{G}_0$ is compact given  the initial value $(\btheta^{(0)}, \bg^{(0)})$.
  \item[(B2)] $F(\btheta, \bg)$ is closed over $\mathcal{G}_0\backslash \mathcal{G}^{*}$, where $\mathcal{G}^{*}=\{\mbox{set of local maxima in the interior}$ $\mbox{ of }~\mathcal{G}\}$. %  i.e., the difference of two sets,
\end{description}

The proof includes five steps as follows:

{\noindent\bf Step 1:} Show that $F(\btheta, \bg)$ is a point-to-point mapping function, a special case of the point-to-set mapping. By the explicit iterative equations (6)--(9) and (10)-(13) in the main text, we know that $F(\btheta, \bg)$ consists of the deterministic combinations and compositions of a series of elementary functions of $(\btheta, \bg)$. Thus, a unique iterative value of $(\btheta^{(t+1)}, \bg^{(t+1)})$ can be obtained. That is, there exists a unique $(\btheta^{(t+1)}, \bg^{(t+1)})$ such that
$(\btheta^{(t+1)}, \bg^{(t+1)}) = F(\btheta^{(t)}, \bg^{(t)})$. Therefore, $F(\btheta, \bg)$ is a point-to-point mapping function.

{\noindent\bf Step 2:} Show that $ELBO(\btheta,\bg)$ is nondecreasing with respect to the sequence $\{(\btheta^{(t)},\bg^{(t)})\}$, i.e.
 $$ELBO((\btheta^{(t-1)},\bg^{(t-1)})) \leq ELBO((\btheta^{(t)},\bg^{(t)})),$$
where  $\btheta^{(t)}=(\mu_{j}^{(t)},\b_j^{(t)}, \lambda_j^{(t)}, j\leq p, \h_i^{(t)}, i\leq n)$ and $\bg^{(t)} = (\mu_{ij}^{(t)},\sigma^{2,{(t)}}_{ij}, i\leq n, j \in G_2 \cup G_3)$.

Note $\ln P(\X|\Y) + \ln P(\Y|\H)=\sum_i \sum_j  \{\ln (P(x_{ij}|y_{ij}) P(y_{ij}|\h_{i}))\}.$
By Jensen's inequality, for any density function of $y_{ij}$, i.e., $q(y_{ij})$, it holds that
\begin{eqnarray*}
% \nonumber to remove numbering (before each equation)
  && \ln (P(x_{ij}|\h_i)) \\
  &=& \ln \int \frac{P(x_{ij}|y_{ij})P(y_{ij}|\h_{i})q(y_{ij})}{q(y_{ij})}dy_{ij} \\
   &\geq &  \mathrm{E}_{q(y_{ij})} \left\{(\ln (P(x_{ij}|y_{ij})P(y_{ij}|\h_{i}   )/g_{ij}(\btheta)) - \ln q(y_{ij})\right\} + \ln g_{ij}(\btheta) \\
   &=&-KL(q(y_{ij}), P(x_{ij}|y_{ij})P(y_{ij}|\h_{i}   )/g_{ij}(\btheta)) + \ln g_{ij}(\btheta),
\end{eqnarray*}
where $\mathrm{E}_{q(y_{ij})}(F(y_{ij}))$ represents taking the expectation of $F(y_{ij})$  with respect to the random variable $y_{ij}$ with a density function $q(y_{ij})$, $g_{ij}(\btheta)=P(x_{ij}|\h_i)$ is the normalizing constant to ensure $P(x_{ij}|y_{ij})P(y_{ij}|\h_{i}   )/g_{ij}(\btheta)$ to be a density function, and $KL(\cdot, \cdot)$ is the KL divergence between two density functions.
Fixing  $\btheta=\btheta^{(t-1)}$, we update $(\mu_{ij}^{(t-1)},\sigma^{2,{(t-1)}}_{ij})$ to $(\mu_{ij}^{(t)},\sigma^{2,{(t)}}_{ij})$ by finding the optimal Gaussian approximation to the posterior distribution of $y_{ij}$, as described in Section 3.1 in the main text.
Therefore, we have
$$KL(q(y_{ij};  \mu_{ij}^{(t-1)},\sigma^{2,{(t-1)}}_{ij}), \frac{P(x_{ij}|y_{ij})P(y_{ij}|\h_{i})}{g_{ij}(\btheta)}) \geq KL(q(y_{ij};  \mu_{ij}^{(t)},\sigma^{2,{(t)}}_{ij}),  \frac{P(x_{ij}|y_{ij})P(y_{ij}|\h_{i})}{g_{ij}(\btheta)}).$$
Furthermore, we have
\begin{eqnarray}
% \nonumber to remove numbering (before each equation)
   && \mathrm{E}_{q(y_{ij}; \mu_{ij}^{(t-1)},\sigma^{2,{(t-1)}}_{ij})}(\ln P(x_{ij}|y_{ij}) + \ln P(y_{ij}|\h_{i})) - \ln q(y_{ij}; \mu_{ij}^{(t-1)},\sigma^{2,{(t-1)}}_{ij}) \nonumber\\
  &\leq &  \mathrm{E}_{q(y_{ij}; \mu_{ij}^{(t)},\sigma^{2,{(t)}}_{ij})}(\ln P(x_{ij}|y_{ij}) + \ln P(y_{ij}|\h_{i}))- \ln q(y_{ij}; \mu_{ij}^{(t)},\sigma^{2,{(t)}}_{ij}), \label{eq:Laplace}
\end{eqnarray}
where $\mathrm{E}_{q(y_{ij}; \mu_{ij}^{(t)},\sigma^{2,{(t)}}_{ij})}F(y_{ij})$ represents taking the expectation of $F(y_{ij})$  with respect to the random variable $y_{ij}$ {which follows} a normal distribution with mean $\mu_{ij}^{(t)}$ and variance $\sigma^{2,{(t)}}_{ij}$. By taking summation of the both sides of equation \eqref{eq:Laplace} over $i=1,\cdots, n$ and $j \in G_2 \cup G_3$, we obtain
\begin{equation}\label{eq:muSigma}
  ELBO(\btheta^{(t-1)},\bg^{(t-1)}) \leq ELBO(\btheta^{(t-1)},(\mu_{ij}^{(t)},\sigma^{2,{(t)}}_{ij}, i\leq n, j\leq p)).
\end{equation}

By the block coordinate maximization principle and the equations (8)--(11) in the main text, we obtain
\begin{equation}\label{eq:theta}
 ELBO(\btheta^{(t-1)}, \bg^{(t)})\leq ELBO(\btheta^{(t)}, \bg^{(t)}).
\end{equation}
Combing with \eqref{eq:muSigma}--\eqref{eq:theta}, we conclude the desired result, i.e.,
$$ELBO(\btheta^{(t-1)},\bg^{(t-1)}) \leq ELBO(\btheta^{(t)}, \bg^{(t)}).$$

{\noindent\bf Step 3.} We show that if $(\btheta_1, \bg_1)\notin \mathcal{G}^{*}$, then $ELBO(\btheta_1, \bg_1) < ELBO(\btheta_2, \bg_2)$ for $(\btheta_2, \bg_2)=F(\btheta_1, \bg_1)$; and if $(\btheta_1, \bg_1)\in \mathcal{G}^{*}$, then
$ELBO(\btheta_1, \bg_1) \leq ELBO(\btheta_2, \bg_2)$  for$(\btheta_2, \bg_2)=F(\btheta_1, \bg_1)$. By the  definition of $\mathcal{G}^{*}$, this result is a direct consequence of the proof of Step 2.

{\noindent\bf Step 4.} We show that $ELBO(\btheta^{(t)}, \bg^{(t)})$ converges monotonically to $L^{*}=ELBO(\btheta^{*}, \bg^{*})$ for some $(\btheta^{*}, \bg^{*}) \in \mathcal{G}^{*}$.

By Assumption (B1), for $(\btheta^{(t)}, \bg^{(t)})\subset \mathcal{G}_0$, we could find a convergent subsequence $(\btheta^{(t_i)}, \bg^{(t_i)})$ converging to the limit $(\btheta^{*}, \bg^{*})$, where $\{t_i: t_i< t_{i+1}, i=1,2,\cdots\}$. Since $ELBO(\btheta, \bg)$ is continuous, it induces that
 $$\lim_{i\rightarrow\infty} ELBO(\btheta^{(t_i)}, \bg^{(t_i)})=ELBO(\btheta^{*}, \bg^{*})$$
 and
 $$ ELBO(\btheta^{(t_i)}, \bg^{(t_i)})\leq ELBO(\btheta^{*}, \bg^{*}),$$
 with $ELBO(\btheta^{(t_i)}, \bg^{(t_i)})$ being monotonic with respect to $i$. Moreover, we have
 $$ELBO(\btheta^{(t)}, \bg^{(t)})\leq ELBO(\btheta^{(t+1)}, \bg^{(t+1)}), t=1,2, \cdots.$$
Then for $t=1,2, \cdots$, there exists a $t_i$ satisfying $t<t_i$ and $ELBO(\btheta^{(t)}, \bg^{(t)}) \leq ELBO(\btheta^{(t_i)},$ $\bg^{(t_i)}) \leq ELBO(\btheta^{*}, \bg^{*})$. That is, $ELBO(\btheta^{(t)}, \bg^{(t)})$ is a monotonic and bounded sequence. Since its subsequence $\{ELBO(\btheta^{(t_i)}, \bg^{(t_i)})\}$ converges to $ELBO(\btheta^{*}, \bg^{*})$, then
\begin{equation}\label{eq:limit}
  \lim_{t\rightarrow\infty} ELBO(\btheta^{(t)}, \bg^{(t)})=ELBO(\btheta^{*}, \bg^{*}).
\end{equation}

{\noindent\bf Step 5.} Finally, we use contradiction approach to show $(\btheta^{*}, \bg^{*}) \in \mathcal{G}^{*}$.

Assume that $(\btheta^{*}, \bg^{*})$ is not in $\mathcal{G}^{*}$. We investigate the subsequence $(\btheta^{(t_i+1)}, \bg^{(t_i+1)})$, where $t_i$ is {(the)} same as the above one in Step 4. Since all members of this sequence are  contained in a compact set, there is a convergent subsequence $\{(\btheta^{(t_{i_k}+1)}, \bg^{(t_{i_k}+1)}), i_1<i_2<\cdots\}$ such that
$$\lim_{k\rightarrow \infty} (\btheta^{(t_{i_k}+1)}, \bg^{(t_{i_k}+1)}) = (\btheta^{**}, \bg^{**}).$$
Note that $\{(\btheta^{(t_{i_k})}, \bg^{(t_{i_k})})\}$ is a subsequence of $\{(\btheta^{(t_{i})}, \bg^{(t_{i})})\}$, we have
$$\lim_{k\rightarrow \infty} (\btheta^{(t_{i_k})}, \bg^{(t_{i_k})}) = (\btheta^{**}, \bg^{**}).$$
Using the monotonicity of $ELBO(\btheta^{(t)}, \bg^{(t)})$ with respect to $t$ and $t_{i_k}<t_{i_k}+1 \leq t_{i_{k+1}}$, we obtain
$$ELBO(\btheta^{(t_{i_{k}})}, \bg^{(t_{i_{k}})}) \leq ELBO(\btheta^{(t_{i_{k}}+1)}, \bg^{(t_{i_{k}}+1)})\leq ELBO(\btheta^{(t_{i_{k+1}})}, \bg^{(t_{i_{k+1}})}),$$
which implies $ELBO(\btheta^{*}, \bg^{*})=ELBO(\btheta^{**}, \bg^{**})$ by the continuity of $ELBO(\cdot, \cdot)$.

Given Condition (B2) and the fact that $(\btheta^{(t_{i_{k}}+1)}, \bg^{(t_{i_{k}}+1)})=F(\btheta^{(t_{i_{k}})}, \bg^{(t_{i_{k}})})$, we can conclude that  $(\btheta^{**}, \bg^{**})=F(\btheta^{*}, \bg^{*})$. However, if $(\btheta^{*}, \bg^{*})$ is not in $\mathcal{G}_{*}$, then according to the results of Step 3, we have  $ELBO(\btheta^{*}, \bg^{*})<ELBO(\btheta^{**}, \bg^{**})$. This leads to a contraction, which implies that  $(\btheta^{*}, \bg^{*})$ must belong to $\mathcal{G}^{*}$.

Combining the results of Step 4 and 5, we  complete the proofs of Theorem 3.1.

\subsection{Details of the algorithm implementation}
In order to enhance the clarity of the practical implementation of the proposed algorithm, we  include additional details of the algorithm implementation. 

Initialization is important for both good estimation performance of parameters and convergence speed of EM algorithm. To obtain a good initial, we utilize the following initialization schema. Specifically, we construct a surrogate matrix $\tilde \X=(\tilde x_{ij})\in \mathbb{R}^{n\times p}$, where $\tilde x_{ij}=x_{ij}$ if $j\in G_1 \cup G_3$ (for continuous and binomial variables) and $\tilde x_{ij}=ln(1+x_{ij})$ if $j\in G_2$ (for the count variables). Then we obtain $\mu_j^{(0)}=\frac{1}{n}\sum_{i=1}^n \tilde x_{ij}$. Next, we perform rank-q SVD decomposition for $\check{\X}=(\tilde x_{ij}-\mu_j^{(0)})\in \mathbb{R}^{n\times p}$, i.e., $\check{\X}=UDV^{\trans}$, where $U\in \mathbb{R}^{n\times q}, D \in \mathbb{R}^{q\times q}$ and $V \times \mathbb{R}^{p\times q}$. Let $\H^{(0)}=\sqrt{n} U$ and $\B^{(0)}=VD/\sqrt{n}$. Then, we initialize the variational parameters as $\tau_{ij}^{(0)}=\tilde x_{ij}$ and $\sigma_{ij}^{2,(0)}=1$. Using this initialization schema, we observed that the variational EM algorithm converges with $|ELBO_t - ELBO_{t-1}|/ |ELBO_{t-1}|<1e-4$ within around 30 iterations and achieves good estimation performance. 
We have obtained relatively good estimation using the above initialization, thus, we do not use other special strategy for coping with multiple local optima.

In addition, the update expression formulation provides a programming advantage by leveraging the efficient C and C++ code in RcppArmadillo, seamlessly linked through Rcpp to enhance computational speed. In the proposed OverGFM algorithm, as outlined in equations (6)-(11), {the update of all parameters} can be executed through matrix operations, offering a programming-friendly approach. Consequently, we utilize the efficient C and C++ code in RcppArmadillo linked through Rcpp to further accelerate the computation.

\section{Additional numerical results}

\subsection{Simulation study}
\noindent\underline{Additional results in Scenario 1}. 
To investigate how the proposed method would fare if overdispersion mechanism is quite different from our assumption, we generate a setting with the overdisperion term from a highly heavy distribution, i.e., t distribution with the degree of freedom 1. Specifically, we generate $(\varepsilon_{i1}, \cdots, \varepsilon_{ip})$ from the multivariate t distribution with mean vector zeros and scale matrix $\sigma^2 \I_p$, i.e., $(\varepsilon_{i1}, \cdots, \varepsilon_{ip}) \sim T(\bbo, \sigma^2\I_p, 1)$, and  vary the overdispersion parameter $\sigma^2$ within the grid $\{0.2,0.3,0.4,0.5\}$. Moreover, we set $(n,p)=(500,500)$ and $(\rho_1, \rho_2, \rho_3) = (4.0, 0.8, 2.4)$. Figure S1 shows that OverGFM still surpasses other methods in terms of estimation accuracy for factor and loading-intercept matrices, while GFM and MRRR  break down with error produced, and LFM produce estimation accuracy near zero. The result from this setting shows OverGFM is not only flexible to the overdispersion but also robust to the heavy-tail data and model misspecification, making it
a highly attractive and favorable choice in practical applications.
\begin{figure}[H]
\centering
\includegraphics[width=0.8\textwidth]{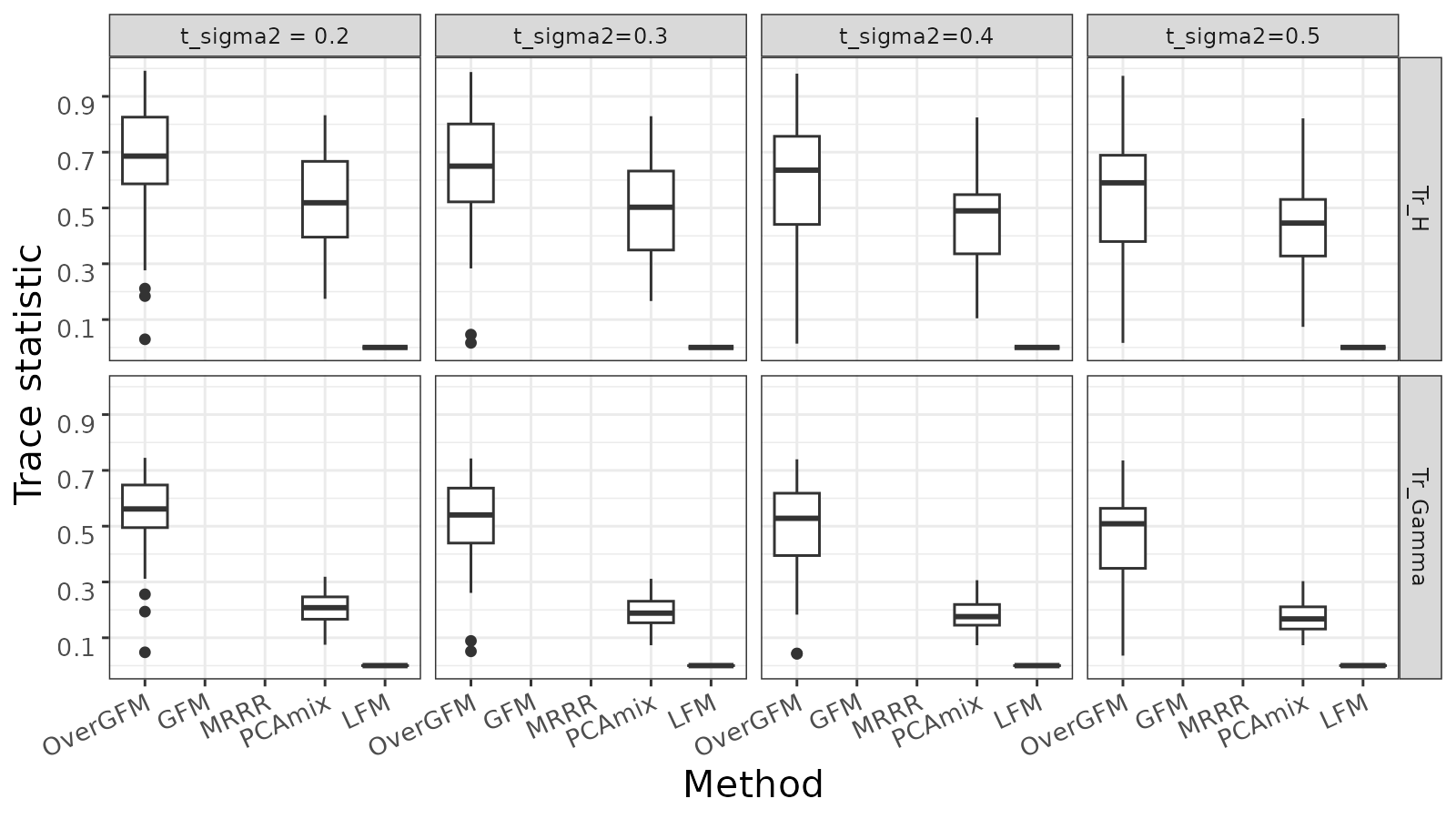}
  \caption{Comparison of estimation accuracy among OverGFM and other methods under overdispersion from a $t$ distribution (degree of freedom is 1) with three mixed-type variables, where $(n,p)=(500,500)$, $q = 6$, $\sigma^2 \in \{0.2,0.3,0.4,0.5\}$, Tr\_H and Tr\_Gamma denote the trace statistics with respect to $\H$ and $\Upsilon$, respectively. Note that the algorithms of both GFM and MRRR break down for all four cases.}\label{fig:heavy_tailed_distribution}
\end{figure}
\noindent\underline{Scenario 7}. 
In this scenario, we explore six distinct cases, encompassing both mixed-type data with two variable types (continuous and count variables, continuous and binary variables, and count and binary variables) and single-type data (continuous variables, count variables, and binary variables). We consider cases involving a single variable type, primarily to compare OverGFM with methods (4), (6), and (7) as detailed in the main text. For each case, we fix the values of $(n, p)$ at (300, 300), the overdispersion parameter $(\sigma^2)$ at 1, the number of factors $(q)$ at 6, and refer to the signal strength settings provided in Table \ref{tab:singlestr}.The data generating process for other parameters is the same as that in Scenario 1.
\begin{table}[!ht]
    \centering\renewcommand\tabcolsep{3.0pt} % 调整表格列间的宽度
    \caption{Simulation settings in the scenario 4: signal strength for each variable type across six cases.}\label{tab:singlestr}
    \begin{tabular}{|c|c|c|c|c|c|c|}
    \hline
        Case & Normal+Poisson & Normal+Binary & Poisson+Binary & Normal & Poisson & Binary \\ \hline
        $\rho_1$ & 0.3 & 0.6 & - & 0.3 & - & - \\ \hline
        $\rho_2$& - & 0.1 & 0.1 & - & - & 0.6 \\ \hline
        $\rho_3$ & 0.4 & - & 0.5 & - & 0.6 & - \\ \hline
    \end{tabular}
\end{table}

We further conduct a comprehensive performance assessment of OverGFM by generating datasets comprising a mixture of two variable types or only single variable type, including continuous, count or binary variables. For the cases with two types, we compare the performance of OverGFM with other methods such as GFM, MRRR and PCAmix, and LFM.
In the case of count data, we compare OverGFM with GFM, MRRR, PCAmix, generalizedPCA, PoissonPCA, PLNPCA, and LFM. For datasets containing binary or continuous variables, we compare OverGFM with GFM, MRRR, PCAmix, generalizedPCA, and LFM. Through these comparisons, we aim to evaluate the effectiveness and efficiency of OverGFM in various data settings. 

\begin{figure}[H]
\centering
\subfigure[Mix of two variable types]{
\includegraphics[width=14cm]{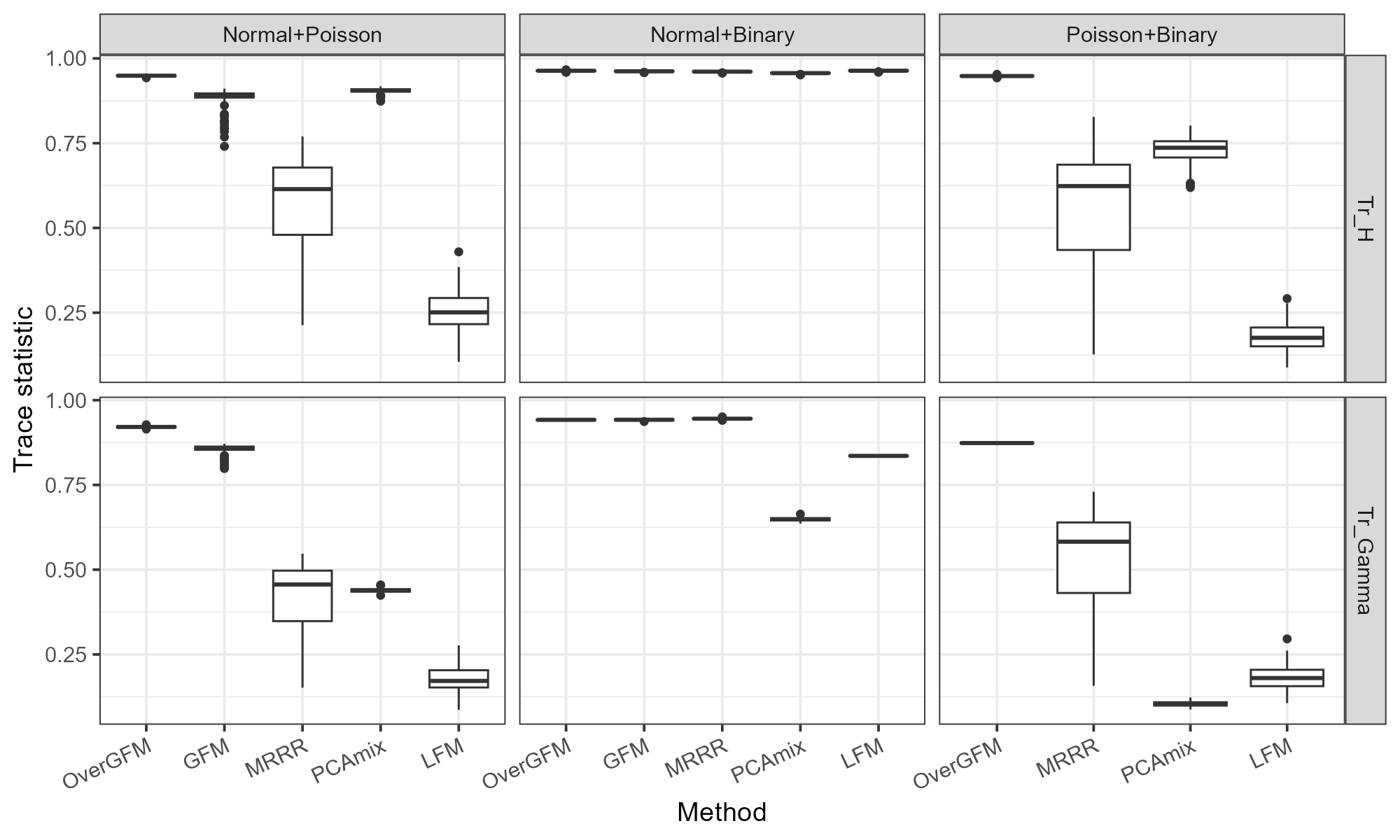}
\label{fig:difn}
}
\subfigure[Only one variable type]{
\includegraphics[width=14cm]{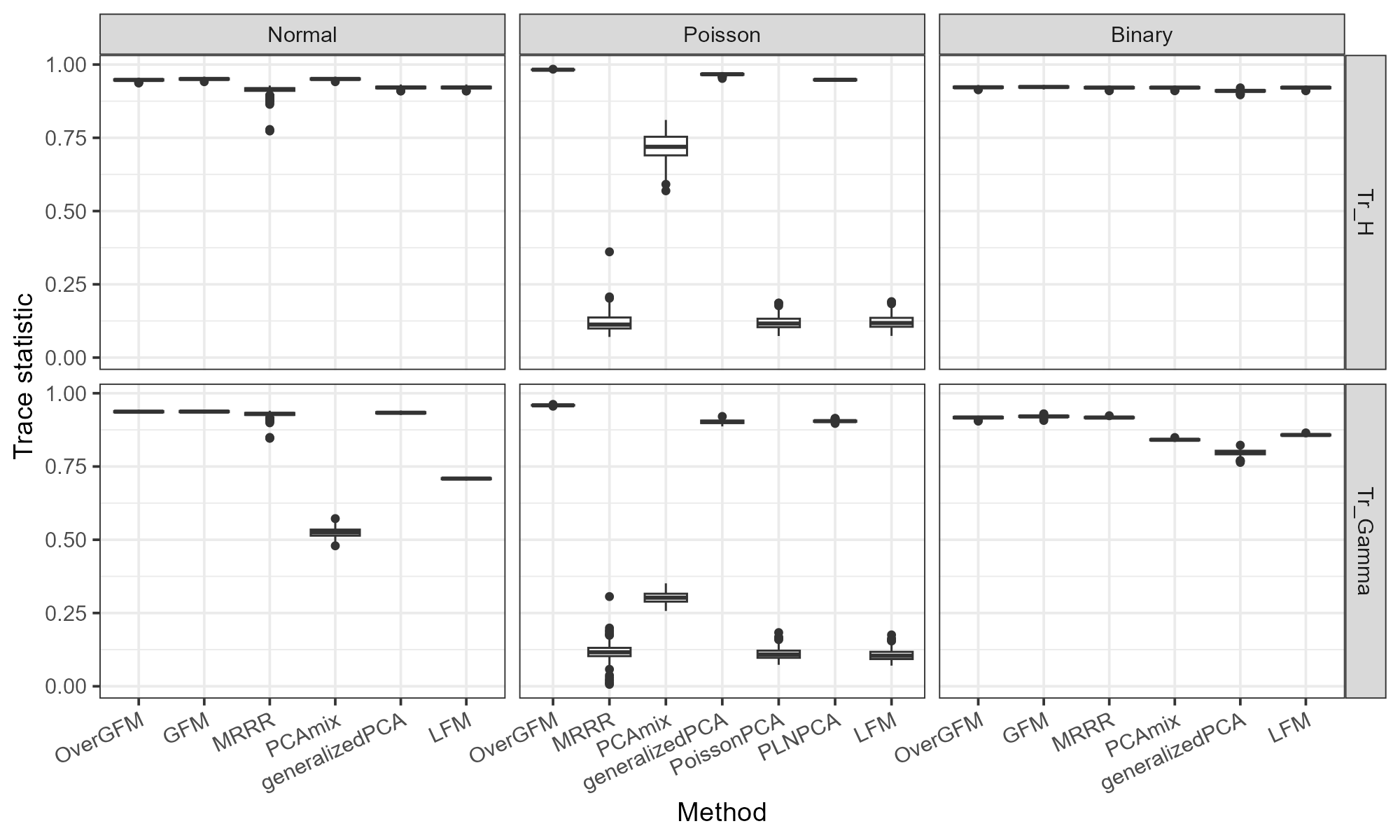}
\label{fig:difp}
}
\caption{Comparison of  estimation accuracy among  OverGFM and other methods under the special cases of mixed-type data, $(n,p)=(300,300)$ and $q=6$. (a): mix of two variable types within normal, Poisson and binary. (b): single variable type within  these three variable types. Note that  GFM does not produce valid results since its algorithm is unstable and breaks down in the case with mix of Poisson and binary variables and the case with only Poisson variables.}\label{fig:specialcases}
\end{figure}

Figure \ref{fig:specialcases} summarizes the estimation accuracy of factor and loading-intercept matrices for OverGFM by comparison with competitors.  Based on our observations, OverGFM consistently demonstrates superior performance compared to other methods in scenarios involving mixed-type data, with one exception being the combination of normal and binary variables. This finding aligns with the conclusions drawn from Scenarios 1-3.
In scenarios involving single-type data, OverGFM outperforms other methods when the variables follow a Poisson distribution. Additionally, when the variables follow a normal or binary distribution, OverGFM either outperforms or performs comparably to GFM, MRRR, and generalizedPCA. In addtion, we observe that GFM is unstale in analyzing the data with Poisson variables and its algorithm breaks down in the case with mix of Poisson and binary variables and the case
with only Poisson variables. These numerical findings strongly suggest that OverGFM is a preferable choice for handling both mixed-type and single-type data compared to methods that can only manage single-type data or existing methods designed for mixed-type data. Overall, our results indicate that OverGFM exhibits superior performance across a range of data scenarios, making it a robust and reliable method for data analysis.

\noindent\underline{Scenario 8}. From the perspective of model settings, GFM is a special case of OverGFM when the overdispersion variance parameters in OverGFM are zeros. To numerically validate this, we delve into the interconnection between OverGFM and GFM. We consider two simple cases involving data from  Gaussian and Poisson distributions, respectively. In Poisson case, we set $\rho_1=0.2$ for Gaussian case, and $\rho_2=0.3$ for Poisson case. We set $(n, p)$ at (300, 300) and other parameters as the same as that in Scenario 1. By varying $\sigma^2$ from 0 to 1, we compare the performance of OverGFM and GFM. Table \ref{tab:two} showed that when the overdispersion variance parameters are zeros ($\sigma^2=0$), GFM indeedly outperforms OverGFM in estimation accuracy. Specifically, in the Gaussian case, although GFM and OverGFM achieve the same average estimation accuracy while GFM has smaller variance; in the Poisson case, GFM has higher estimation accuracy in both average and variance of trace statistics of factor score matrix and loading-intercept matrix. This is accordance with the intuition since GFM is the true model. In contrast, OverGFM pays the price in estimating the redandant overdispersion parameters in this case. As the overdispersion parameters increase, we observed that OverGFM outperforms GFM, which means the model misspecification of GFM causes the loss of estimation performance. 
\begin{table}[H]
    \centering\caption{Comparison of estimation accuracy among OverGFM and GFM  under Gaussian and Poisson cases, where $(n,p) = (300,300), q = 6, \sigma^2 \in \{0,0.5, 1\}$. Reported are the average (standard deviation in parathesis) for performance metrics, where Tr\_H and Tr\_Gamma denote the trace statistics with respect to $\H$ and $\Upsilon$, respectively.}\label{tab:two}
    \begin{tabular}{llllll}
    \hline
         & Metric & Gaussian case & ~ & Poisson case & ~ \\ \hline
        ~ & ~ & OverGFM & GFM & OverGFM & GFM \\ \hline
        $\sigma^2=0$ & Tr\_H & 1.00 (3.4e-16) & 1.00(3.1e-16) & 0.974(9.3e-4) & 0.975(8.7e-4) \\ 
        ~ & Tr\_Gamma & 1.00 (2.2e-16) & 1.00(2.2e-16) & 0.944(1.9e-3) & 0.971(1.3e-3) \\ \hline
        $\sigma^2=0.5$ &  Tr\_H & 0.934(1.6e-3) & 0.945(1.7e-3) & 0.952(1.8e-3) & 0.938(2.4e-3) \\ 
        ~ & Tr\_Gamma & 0.945(2.4e-3) & 0.945(2.4e-3) & 0.917(2.2e-3) & 0.897(2.5e-3) \\ \hline
        $\sigma^2= 1$ &  Tr\_H & 0.901(4.4e-3) & 0.897(7.0e-3) & 0.931(2.4e-3) & 0.883(4.7e-3) \\ 
        ~ & Tr\_Gamma & 0.896(4.1e-3) & 0.894(4.2e-3) & 0.875(2.9e-3) & 0.782(4.3e-3) \\ \hline
    \end{tabular}
\end{table}

In Scenario 7, we observe that GFM performs not stable for the cases with Poisson variables.  To gain a deeper understanding of these  results, we  further examine the numerical performance of GFM, comparing it with OverGFM in Poisson cases with overdispersion variance parameters being zeros.  Specifically, we change the magnitude of count values by setting different $\rho_2$'s that control the signal strength and magnitude of count values in the Poisson case. The results are consolidated in Table \ref{tab:instablePois}. When $\rho_2=0.3$ and the maximum count value is $max=483$, the GFM algorithm operates smoothly and outperforms OverGFM in estimating both the factor score matrix and loading-intercept matrix. However, as the maximum count values increase to 3155, 77727, and 319516 for $\rho_2$ values of 0.5, 0.7, and 0.8 respectively, the GFM algorithm exhibits instability, resulting in failure of some runs. Additionally, in these scenarios, GFM performs inferior to OverGFM in estimating the factor score matrix. Therefore, the numerical performance of GFM is adversely affected by the instability of its designed algorithm.
\begin{table}[H]
    \centering\caption{Comparison of estimation accuracy among OverGFM and GFM  under Poisson cases, where $(n,p) = (300,300), q = 6, \sigma^2 =0$, and $\rho_2$ is varied from 0.3 to 0.8 to generate different magnitude of count values. Reported are the average performance metrics (with standard deviation in parentheses) and the number of breakdowns (Failure times) encountered by the corresponding algorithms, where $max$ is the maximum of the count values over 200 runs, Tr\_H and Tr\_Gamma denote the trace statistics with respect to $\H$ and $\Upsilon$, respectively.}\label{tab:instablePois}
    \begin{tabular}{llll}
    \hline
        ~ & Metrics & OverGFM & GFM \\ \hline
        $\rho_2=0.3$ & Tr\_H  & 0.974(9.3e-4) & 0.975(8.7e-4) \\ 
        max=483 & Tr\_Gamma & 0.944(1.9e-3) & 0.971(1.3e-3) \\
        ~ & Failure times & 0 & 0 \\ \hline
        $\rho_2=0.5$ &Tr\_H  & 0.993(2.8e-4) & 0.988(2.9e-2) \\ 
        max=3155 & Tr\_Gamma & 0.966(1.1e-3) & 0.987(1.5e-2) \\ 
        ~ & Failure times & 0 & 1 \\ \hline
         $\rho_2=0.7$ & Tr\_H  & 0.997(1.3e-4) & 0.984(3.9e-2) \\ 
        max=77727 & Tr\_Gamma & 0.976(7.4e-4) & 0.983(2.5e-2) \\ 
        ~ & Failure times & 0 & 87 \\ \hline
         $\rho_2=0.8$ &Tr\_H  & 0.998(2.5e-4) & 0.996(1.2e-2) \\ 
        max=319516 & Tr\_Gamma & 0.980(1.9e-3) & 0.993(1.5e-2) \\ 
        ~ & Failure times & 0 & 135 \\ \hline
    \end{tabular}
\end{table}

To further explore why the instability happens in the Poisson case with large count values for GFM, we turn to investigating the details of the algorithm of GFM. We found that, for ease of implementation, GFM relies on existing packages, such as {\it glmfit} function in the MATLAB or {\it glm} function in the R software. This approach translates the updating of the factor score matrix and loading matrix into  $n+p$ generalized linear model (GLM) fittings. Given $\H$, the algorithm updates $(\mu_j, \b_j)$ for $j=1, \cdots, p$; then given $\{(\mu_j, \b_j),j=1, \cdots, p\}$, it updates $\h_i$ for $i=1, \cdots, n$. Since both $\H$ and $\{(\mu_j, \b_j),j=1, \cdots, p\}$ are unknown parameters, iteratively applying  {\it glmfit}  or  {\it glm} to fit numerous GLM models becomes vulnerable to instability when the magnitude of count values is high. If any one of the $n+p$ GLM models encounters an exception, it will negatively impact the numerical performance of GFM,  and even lead to the breakdown of the GFM algorithm itself. In contrast, OverGFM possesses its own variational EM algorithm that does not rely on any pre-existing algorithms, rendering it more robust in terms of numerical performance for high-magnitude count values. As evident in Table \ref{tab:instablePois}, OverGFM experiences zero failures. All these findings indicate that OverGFM is a more robust choice compared to GFM, particularly for real-world applications with high noise levels.

\subsection{Real data analysis}
\subsubsection{Other results in scRNA-seq data analysis}
\begin{figure}[H]
\centering
\includegraphics[width=0.8\textwidth]{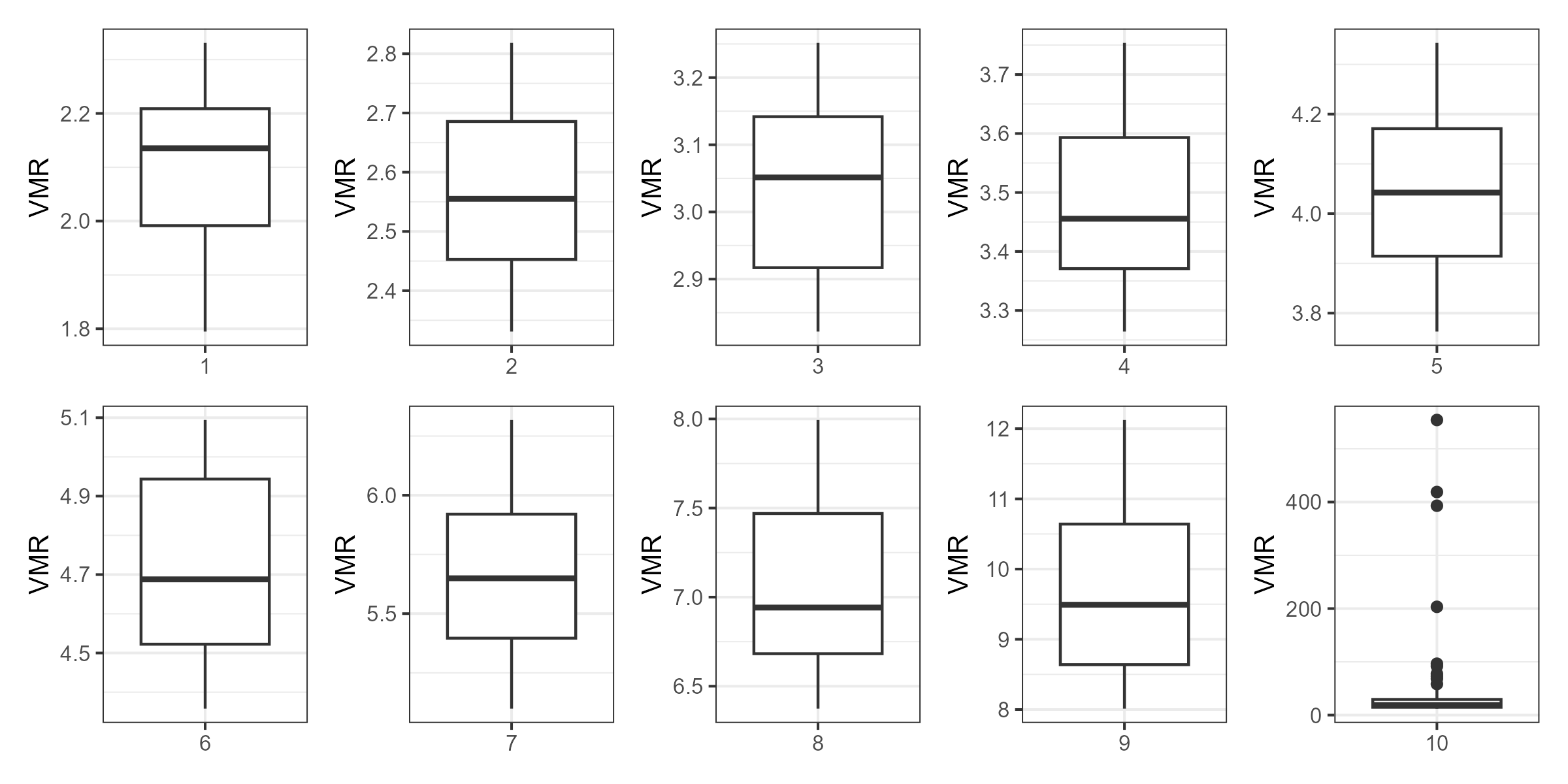}
  \caption{Boxplots illustrating the Variance-to-Mean Ratio (VMR) metric across the ten groups of gene expression levels. A VMR significantly greater than 1 suggests overdispersion.}\label{fig:VMR}
\end{figure}

\begin{figure}[H]
\centering
\includegraphics[width=0.8\textwidth]{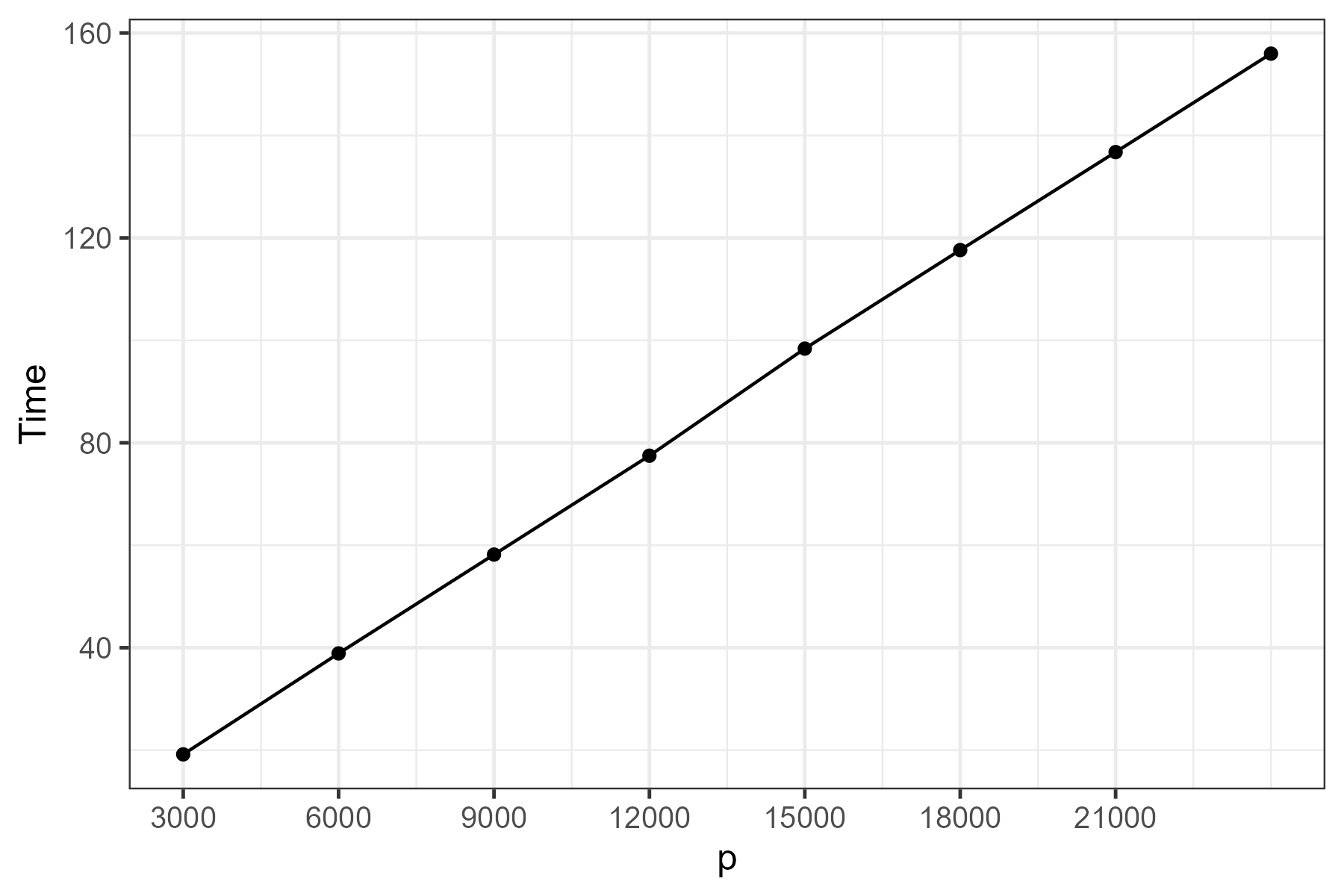}
  \caption{The running time (in seconds) of OverGFM varies as the dimension of variables changes based on the selection of a different number of highly variable genes.}\label{fig:}
\end{figure}

\begin{figure}[H]
\centering
\includegraphics[width=0.8\textwidth]{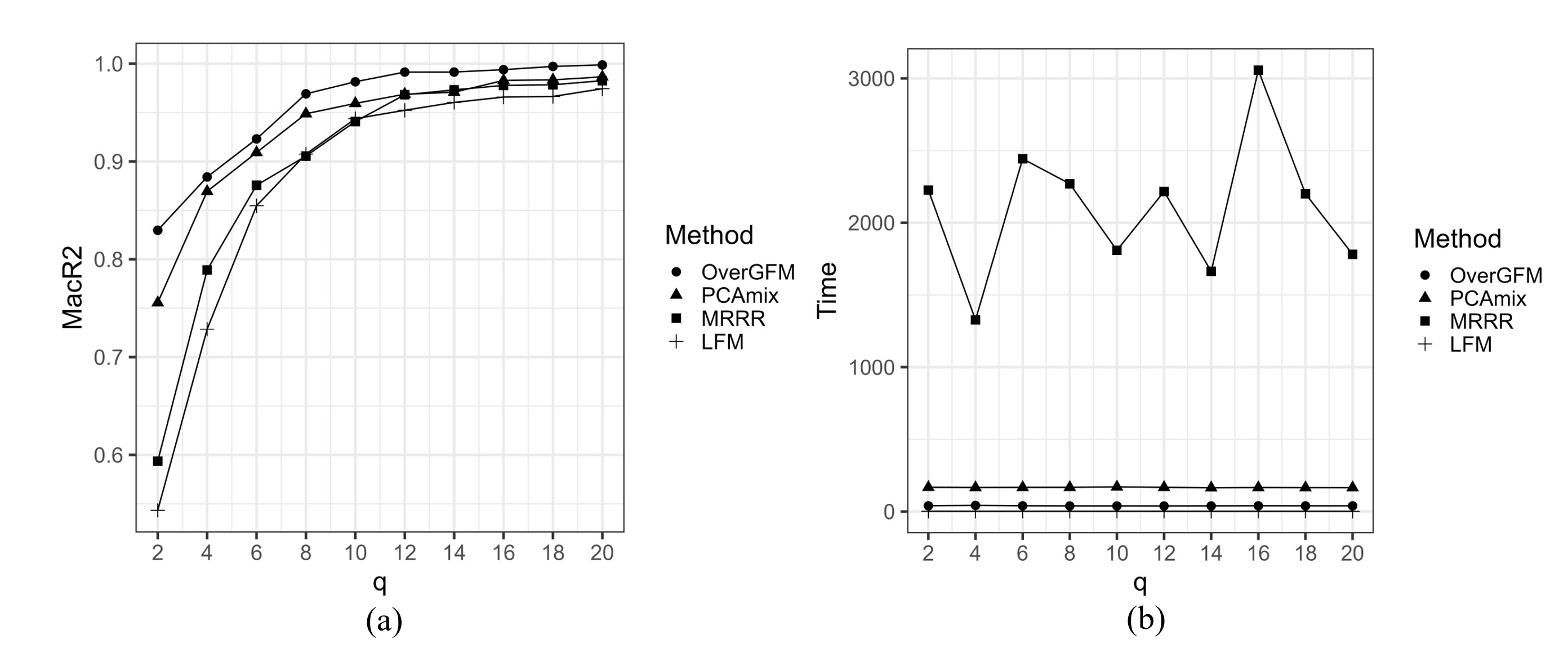}
  \caption{Comparison of OverGFM and three other methods when 2000 highly variable genes is selected.
(a) The adjusted McFadden’s pseudo R2 across different numbers of factors; (b) The running
time (in seconds) across different numbers of factors.}\label{fig:}
\end{figure}

To explore their roles in cellular functional mechanisms, we conducted Gene Ontology (GO) gene set enrichment analysis (GSEA) to investigate the functions of subtypes of Purkinje neurons , respectively. For each subtype from the same cell type, we performed GO GSEA for the marker genes of this subtype, in which the marker genes are defined by the genes with log-fold change greater than 0.25~\citep{liu2023probabilistic} and the adjusted p-value less than 0.05.

By excluding the overlapped GO pathways from each of three categories of functional annotations: biological process (BP), cellular component (CC), and molecular function (MF), Figure \ref{fig:enrichInter} showed the top five GO pathways  from the remaining pathways for the subtypes of Purkinje neurons. We observed the two subtypes pose different functions. For example, Cluster 1 displayed enrichment in various processes in the BP category, including postsynaptic cytoskeleton organization, glucose catabolic process, and cellular response to retinoic acid. In the MF category, Cluster 1 exhibited enrichment in protein serine/threonine kinase activity, protein serine kinase activity, and translation elongation factor activity.
In contrast, Cluster 5 showed enrichment in a different set of BP processes, such as positive regulation of the regulated secretory pathway, dephosphorylation, and response to amine. In the MF category, Cluster 5 was enriched in corticotropin-releasing hormone binding, carboxylic acid binding, and GABA-A receptor activity. These findings indicate that the two subtypes of Purkinje neurons have different functional profiles, as they differ significantly in the enriched GO pathways within each functional annotation category. 

\begin{figure}[H]
\centering\includegraphics[width=13cm]{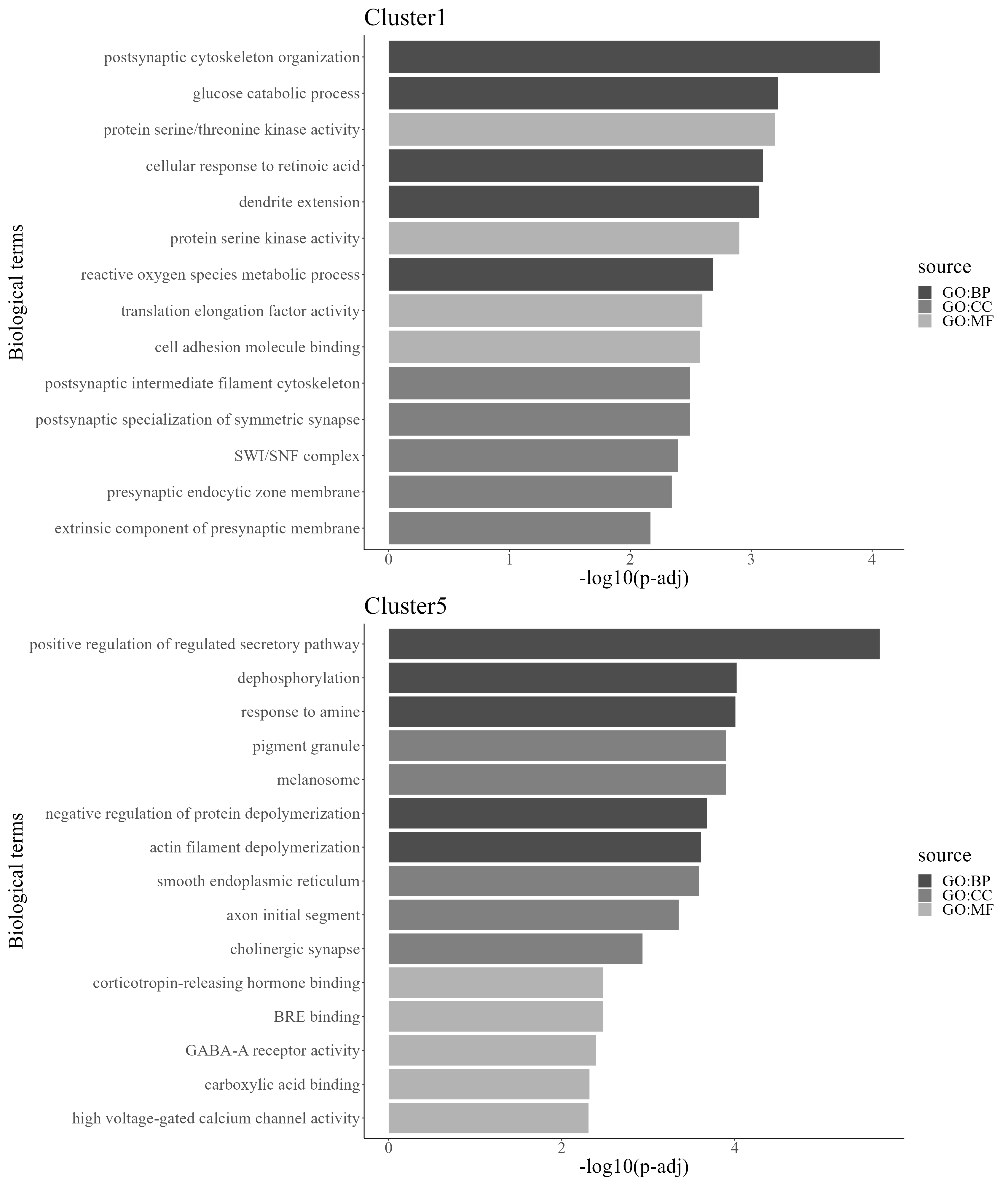}
  \caption{Gene set enrichment analysis for identifying the different functions of the two subtypes of Purkinje neurons, corresponding clusters 1 and 5.}\label{fig:enrichInter}
\end{figure}

\subsubsection{Other results in SNARE-seq data analysis}

\begin{table}[H]
    \centering\scriptsize\renewcommand\tabcolsep{2.5pt}
    \caption{The top 50 genes with the largest loading magnitudes for each of the five directions. In the top genes of loading 3, A83* and A83** represents the genes A830018L16Rik and A830036E02Rik, respectively.}
    \begin{tabular}{|l|l|l|l|l|l|l|l|l|l|l|l|}
    \hline
        Rank & Load1 & Load2 & Load3 & Load4 & Load5 & Rank & Load1 & Load2 & Load3 & Load4 & Load5 \\ \hline
        1 & Il1rapl2 & Galntl6 & Erbb4 & Adarb2 & Adarb2 & 26 & Slit3 & Ncam2 & Pde4d & Gabrg3 & Kcnip4 \\ \hline
        2 & Fam19a1 & Il1rapl2 & Fam19a1 & Erbb4 & Erbb4 & 27 & Rbfox1 & Unc5c & Dlc1 & Epha6 & Trhde \\ \hline
        3 & Pcdh15 & Pcdh15 & Mgat4c & Galntl6 & Galntl6 & 28 & Lrp1b & Cntn4 & Unc5d & Zmat4 & Cadps2 \\ \hline
        4 & Cntn5 & Kcnc2 & Cdh18 & Nxph1 & Nxph1 & 29 & Opcml & Astn2 & Frmpd4 & Dgkb & Rora \\ \hline
        5 & Lrrtm4 & Cntn5 & Hs3st4 & Fam19a1 & Il1rapl2 & 30 & Gm26871 & Fat3 & Garnl3 & Cadps2 & Ncam2 \\ \hline
        6 & Egfem1 & Lsamp & Grip1 & Zfp804b & Fam19a1 & 31 & Agbl4 & Dcc & Pdzrn3 & Ncam2 & Pde1a \\ \hline
        7 & Sgcz & Sgcd & Rorb & Sox2ot & Npas3 & 32 & Chrm3 & Ntm & Rbfox1 & Pde4d & Ptchd4 \\ \hline
        8 & Gm28928 & Sgcz & Zfpm2 & Luzp2 & Gpc5 & 33 & Rgs6 & Frmpd4 & Sema6d & Astn2 & Pde4d \\ \hline
        9 & Gpc6 & Gm28928 & Foxp2 & Npas3 & Grip1 & 34 & Fstl4 & Lrp1b & A83* & Grm1 & Cntn4 \\ \hline
        10 & Hs6st3 & Gpc6 & Gm28928 & Gpc5 & Thsd7a & 35 & Kctd16 & Opcml & Pde7b & Dcc & Kcnh5 \\ \hline
        11 & Sorcs3 & Hs6st3 & Gpc6 & Cdh18 & Pcdh15 & 36 & Kcnb2 & Grik2 & Mctp1 & Frmpd4 & Slc1a2 \\ \hline
        12 & Cntnap5a & Cntnap2 & Hs6st3 & Hs3st4 & Kcnc2 & 37 & Nrxn3 & Agbl4 & Gria1 & Garnl3 & Astn2 \\ \hline
        13 & Epha6 & Cntnap5a & Cntnap5a & Grip1 & Ptprm & 38 & Tenm2 & Kctd16 & Tenm1 & Gria1 & Grm1 \\ \hline
        14 & Lingo2 & Etl4 & Etl4 & Kcnc2 & Lrrtm4 & 39 & Ptprk & Kcnb2 & Fstl4 & Grik2 & Pdzrn4 \\ \hline
        15 & Dgkb & Gabrg3 & Gabrg3 & Kirrel3 & Rorb & 40 & Gria4 & Lrrc4c & Kctd16 & Agbl4 & Slit3 \\ \hline
        16 & Kcnip4 & Sorcs1 & Htr1f & Lrrtm4 & Sgcd & 41 & Ptprd & Nrxn3 & Kcnb2 & Pde4b & Rbfox1 \\ \hline
        17 & Pex5l & Epha6 & Fam19a2 & Rorb & Gm28928 & 42 & Cdh12 & Xkr4 & Slc35f1 & Chsy3 & Gria1 \\ \hline
        18 & Kcnq5 & Lingo2 & Epha6 & Cdh13 & Slc24a3 & 43 & Car10 & Tenm2 & Neto1 & Enox1 & Opcml \\ \hline
        19 & Nkain2 & Kcnip4 & Lingo2 & Zbtb20 & Brinp3 & 44 & Nlgn1 & Gria4 & Ptprk & Cpne6 & Gm26871 \\ \hline
        20 & Pde1a & Pex5l & Grm8 & Lsamp & Nell1 & 45 & Cdh10 & Ptprd & Me3 & Ptprd & Chrm3 \\ \hline
        21 & Pde4d & Kcnq5 & Kcnq5 & Sgcz & Sorcs1 & 46 & 1-Mar & Slc2a13 & A83** & Dock4 & Rgs6 \\ \hline
        22 & Cntn4 & Nkain2 & Nkain2 & Gm28928 & Fam19a2 & 47 & Prkg1 & Nlgn1 & Cdh12 & Cdh12 & Lrrc4c \\ \hline
        23 & Fat3 & Grid2 & Rora & Gpc6 & Robo3 & 48 & Ntrk3 & Cdh10 & Clstn2 & Car10 & Cpne6 \\ \hline
        24 & Ntm & Hcn1 & Pde1a & Brinp3 & Zmat4 & 49 & Cacna2d3 & Cacna2d3 & Prkg1 & Ntrk3 & 1-Mar \\ \hline
        25 & Pdzrn3 & Rora & Prr16 & Cntnap5a & Dgkb & 50 & Slc8a1 & Zfp385b & Cacna2d3 & Slc8a1 & Ntrk3 \\ \hline
    \end{tabular} 
\end{table}

\begin{figure}[H]
\centering\includegraphics[width=13cm]{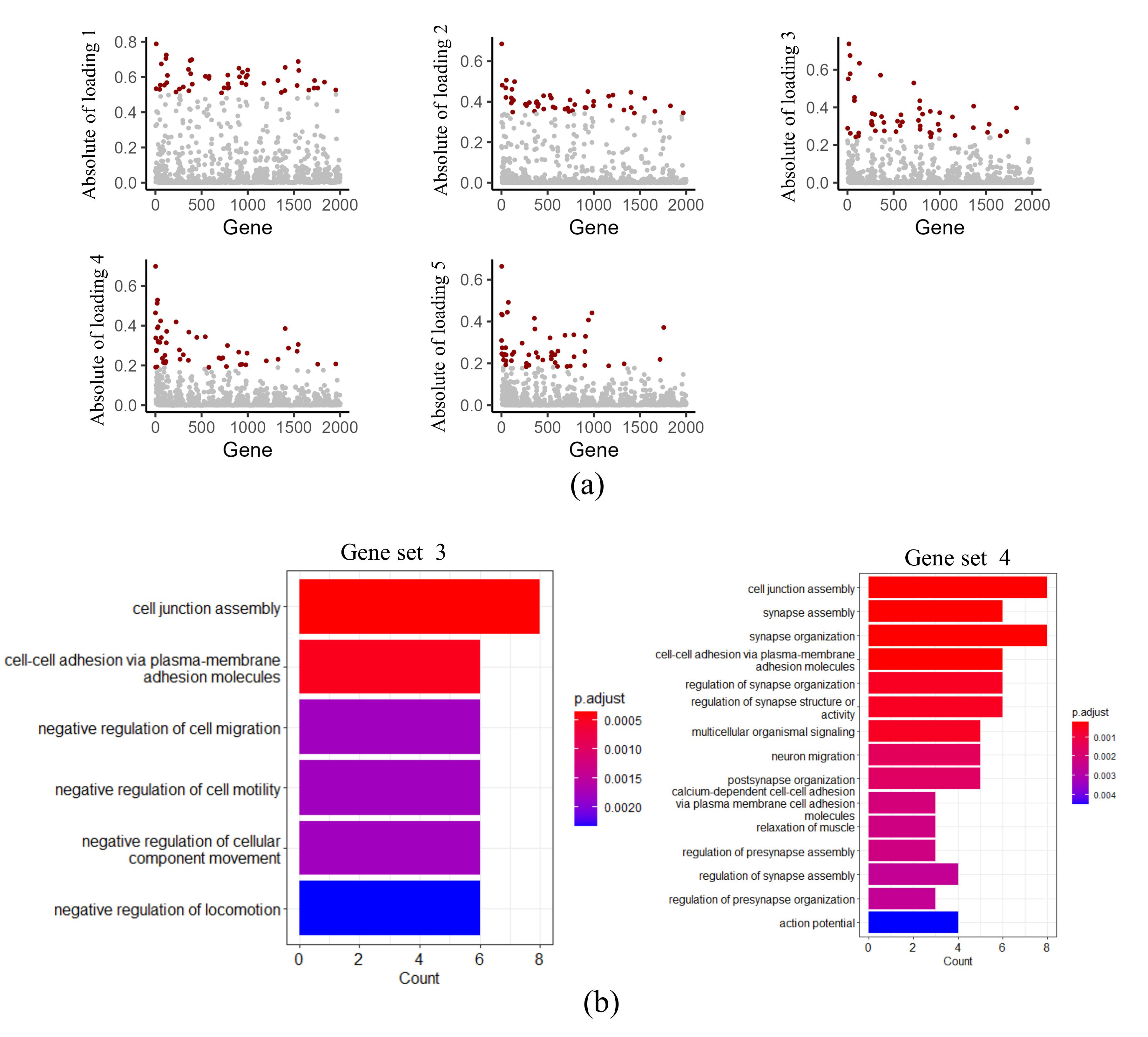}
  \caption{SNARE-seq data analysis using OverGFM. (a) The absolute value of the loadings in each of five directions. The points with red color {represent} the top 50 genes. (b) Barplots showing significant pathways from gene set enrichment analysis for two gene sets (sets 3 and 4) in the biological process category of the GO database. }\label{fig:enrichInter}
\end{figure}

\bibliographystyle{WileyNJD-AMA}%{plain}%

\bibliography{reflib1.bib}